\documentclass[a4paper]{JAC2003}
\usepackage{graphicx}
\usepackage{color}
\usepackage{framed}
\usepackage{comment}
\definecolor{shadecolor}{gray}{0.875}
\definecolor{myred}{rgb}{0.75,0,0}
\definecolor{myblue}{rgb}{0,0.25,0.45}
\definecolor{mygreen}{rgb}{0,0.45,0.25}
\usepackage[toc,page]{appendix}
\usepackage{amsmath,amssymb}
\usepackage{multirow}
\usepackage{graphicx}
\usepackage{array}
\usepackage{alltt}
\usepackage{amssymb}
\usepackage{epstopdf}
\usepackage{wrapfig}
\usepackage{subfigure}
\usepackage{float}
\usepackage{verbatim}
\floatstyle{boxed}
\usepackage{bm}
\usepackage{listings}
\usepackage{booktabs}
\usepackage{varwidth}
\usepackage{xcolor}
\newcommand{\comments}[1]{}
\newcommand{\jx}{J}
\newcommand{\I}{I}
\newcommand{\ho}{h.\,o.~}
\newcommand{\lr}{\rm l.\,r.~}
\newcommand{\re}[1]{(\ref{#1})}
\newcommand{\Fnon}{F_{\rm nonl}}
\newcommand{\eplm}{\mathrm{e}^{i m \phi}}

\newcommand{\no}{\nonumber}
\newcommand{\beq}{\begin{equation}}
\newcommand{\eeq}{\end{equation}}
\newcommand{\beqa}{\begin{eqnarray}}
\newcommand{\eeqa}{\end{eqnarray}}
\newcommand{\beqas}{\begin{eqnarray*}}
\newcommand{\eeqas}{\end{eqnarray*}}

\begin{document}
\newcommand{\appr}{\;$\approx$\;} \newcommand{\en}{n_{\sigma}}
\newcommand{\enxy}{n_x,n_y} \newcommand{\bb}{\lambda}
\newcommand{\ang}{\alpha} \newcommand{\np}{N_{\rm b}}
\newcommand{\nip}{{ N}_{}} \newcommand{\murad}{~\mu{\rm rad}}
\newcommand{\e }[1]{\mathrm{e}^{:#1:}}
\title{ANALYSIS OF LONG-RANGE STUDIES IN
  THE LHC -- COMPARISON WITH THE MODEL} \author{D.~Kaltchev, TRIUMF, Vancouver, Canada,
  W.~Herr, CERN, Geneva, Switzerland}

\maketitle
\begin{abstract}
  We find that the observed dependencies (scaling) of long-range beam--beam effects on the
beam  separation and intensity  are consistent with the simple assumption that,
  all other parameters being the same, the quantity preserved during
  different set-ups is the first-order smear as a function of
  amplitude.
\end{abstract}
\section{Introduction}
\subsection{The Proposed Method}
In several Machine Development (MD) studies
(see Ref.~\cite{mdnote} and the references therein),
reduced crossing angles have been used to enhance long-range beam--beam
effects and thus facilitate their measurement.  The basic assumption
made in this paper is that under such conditions, a single
non-linearity, the one caused by beam--beam, dominates the dynamics.
Hence the method followed: we choose some simple low-order dynamical
quantity that characterizes phase space distortion and assume that
when this quantity is the same, the behaviour of the system is the
same.  A most obvious candidate is the first-order smear -- the
r.m.s.\@ deviation of the phase-space ellipse from the perfect one.  At
a fixed amplitude, smear is defined as the averaged generalized
Courant--Snyder invariant over the angle variable \cite{pac09}.

An analytical expression has previously been found \cite{pac09} for the
smear $S$ as a function of amplitude $\en$.  Suppose that the
parametric dependence of $S(n)$ on several beam--beam related
parameters -- the relativistic $\gamma$, the number of particles per bunch
$N_b$, the crossing angle $\alpha$, and the normalized separations $n_{\rm
  l.r.}$ -- is known. According to the above assumption, for two machine
configurations {\em a} and {\em b} one should have
\begin{equation}
  S(\en; N_b^a , n^a_{\rm l.r.},  \alpha^a,   \gamma^a) =
  S(\en; N_b^b , n^b_{\rm l.r.},  \alpha^b,   \gamma^b) .
  \label{params}
\end{equation}
As a particular application of Eq.~(1), we considered two experiments
where the intensities are $N_b^{a}$ and $N_b^{b}$. All other parameters
being the same, given $\alpha^a$, one can compute the expected
$\alpha^b$.  Our task will be to show that the result agrees with
observations.
\subsection{Analytical Calculation of Invariant and Smear }
Our derivation of $S(\en)$ is based on the Lie algebraic method --
concatenation of Lie-factor maps -- and is valid only to first order in
the beam--beam parameter and in one-dimension, in either the horizontal or
the vertical plane, but for an arbitrary distribution of beam--beam
collisions, head-on or long-range, around the ring.

For a ring with a {\it single head-on collision point}, Hamiltonian
perturbation analysis of the beam--beam interaction without or with
a crossing angle has been done by a number of authors, mostly in the
resonant case. Non-linear invariants of motion, both non-resonant and
resonant, were analysed by Dragt \cite{dragt}, with the one-turn map as
observed immediately after the kick being 
\begin{wrapfigure}{l}{3cm}
  \includegraphics[width=2.7cm,angle=0]{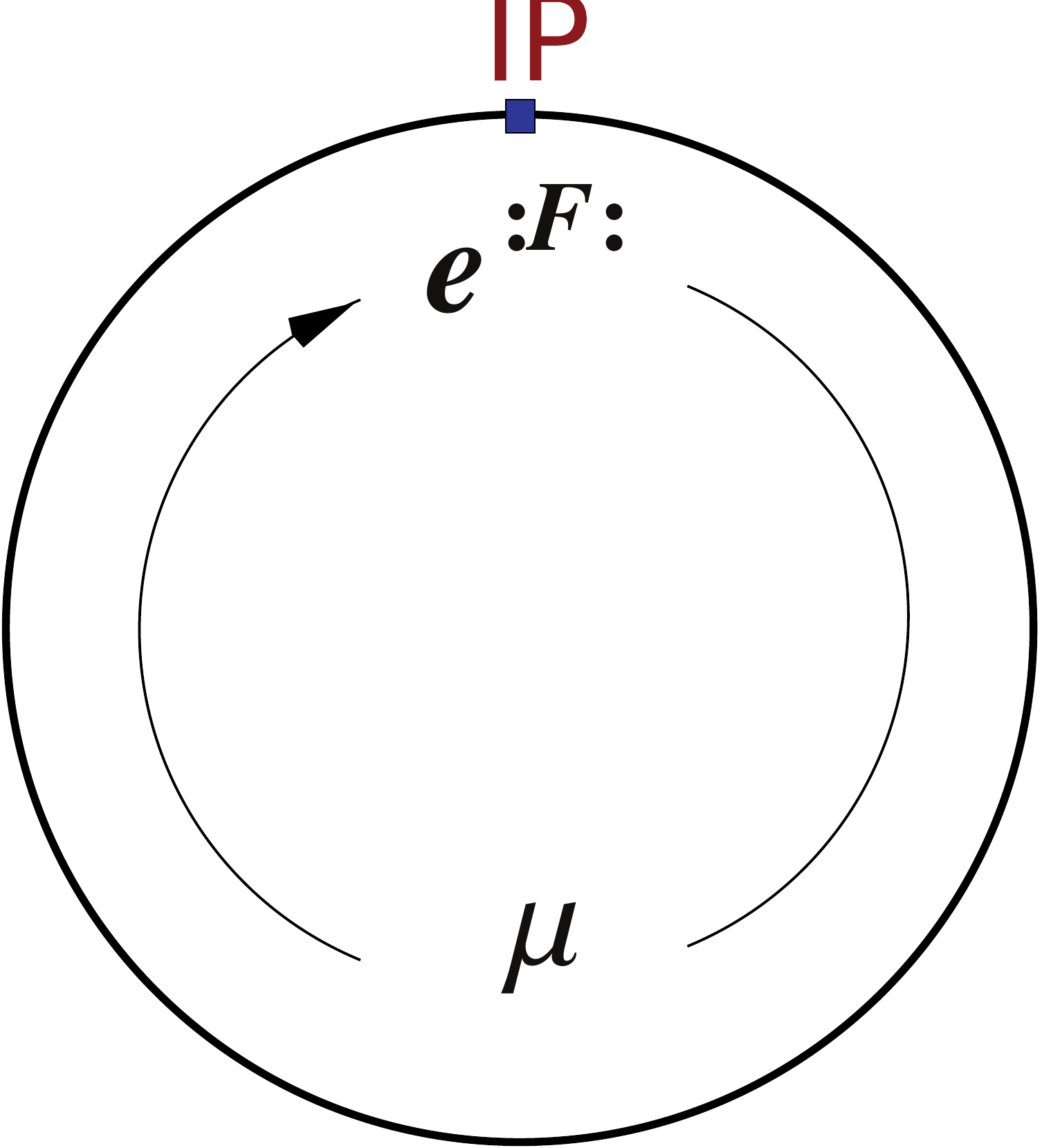}
\end{wrapfigure}
\begin{equation}
  R\; \mathrm{e}^{:{\mathit{F}}:} = \mathrm{e}^{:h:} \,.
  \label{oneip}
\end{equation}
Here, $R=\mathrm{e}^{:f_2:}$ is the linear one-turn map and the kick factor $F$
is the beam--beam potential (or Hamiltonian).  
For small perturbations and far from resonances, particle coordinates in phase
space are restricted on the Poincar\'{e} surface of section
\begin{equation}
  h=\mathrm{const}.
  \label{hconst}
\end{equation}
A detailed derivation of $h$ to first order in the beam--beam
perturbation strength can be found in A.~Chao's lectures:
\begin{eqnarray}
  h(J,\phi) =
  -\mu J+ \sum_{n=-\infty}^{\infty}{c^{\rm (ho)}_n(J) \frac{n\, \mu}{2 \sin \frac{n \mu}{2}} \mathrm{e}^{i n (\phi+\mu/2)}  }\,,
  \label{hafi}
\end{eqnarray}
where $\mu$ is the ring phase advance and $c^{\rm (ho)}_n(J)$ are
coefficients in the Fourier expansion of $F$, when the latter is
rewritten in action-angle coordinates $J,\phi$. The coefficients are
shown to be related to the modified Bessel functions.  Analytical
expressions for the invariant $h$, the first-order smear, and the second-order
detuning for the case of non-linear multipole kicks distributed in an
arbitrary way around the ring have been derived by Irvin and Bengtsson
\cite{irwinsmear}. Smear, the distortion of the ideal phase-space
ellipse, is formally defined in Ref.~\cite{standard}. Finally, note that
extracting the smear is a natural step in the procedure that brings
the map into its normal form \cite{forest}.

In Ref.~\cite{resonances}, following the Lie algebraic procedure in Refs.
\cite{chaolect} and \cite{irwinsmear}, we generalized Eq.~\re{hafi} to
describe multiple head-on kicks (IP1 and IP5) for the case of the LHC. In
Ref.~\cite{pac09}, an expression was presented that was valid for an arbitrary number of
head-on (h.o.) and long-range (l.r.) collisions.  This expression, to
be derived in detail next, has been used on several occasions to
interpret results from SixTrack simulations.
\section{Derivation of the Invariant}
\subsection{ Multiple Collision Points}
The horizontal betatronic motion of a weak-beam test particle depends
on its initial amplitude $\en$ (in units of $\sigma$) and the
collision set: a set of all \ho and \lr collisions, also known as Interaction
Points (IPs), that this particle sees over a single revolution.  Let
us label the set with an index $k$, limiting ourselves to only IPs
located within the main interaction regions IR5 (horizontal crossing)
and IR1 (vertical crossing).  In the case of 50~ns bunch spacing, $k$
ranges from 1 to 34, which includes 32 long-range IPs ($N_{\lr}=32$).

\begin{figure}[h]
  \begin{center}
    \vspace{-.1cm}
    \includegraphics[width=.5\textwidth]{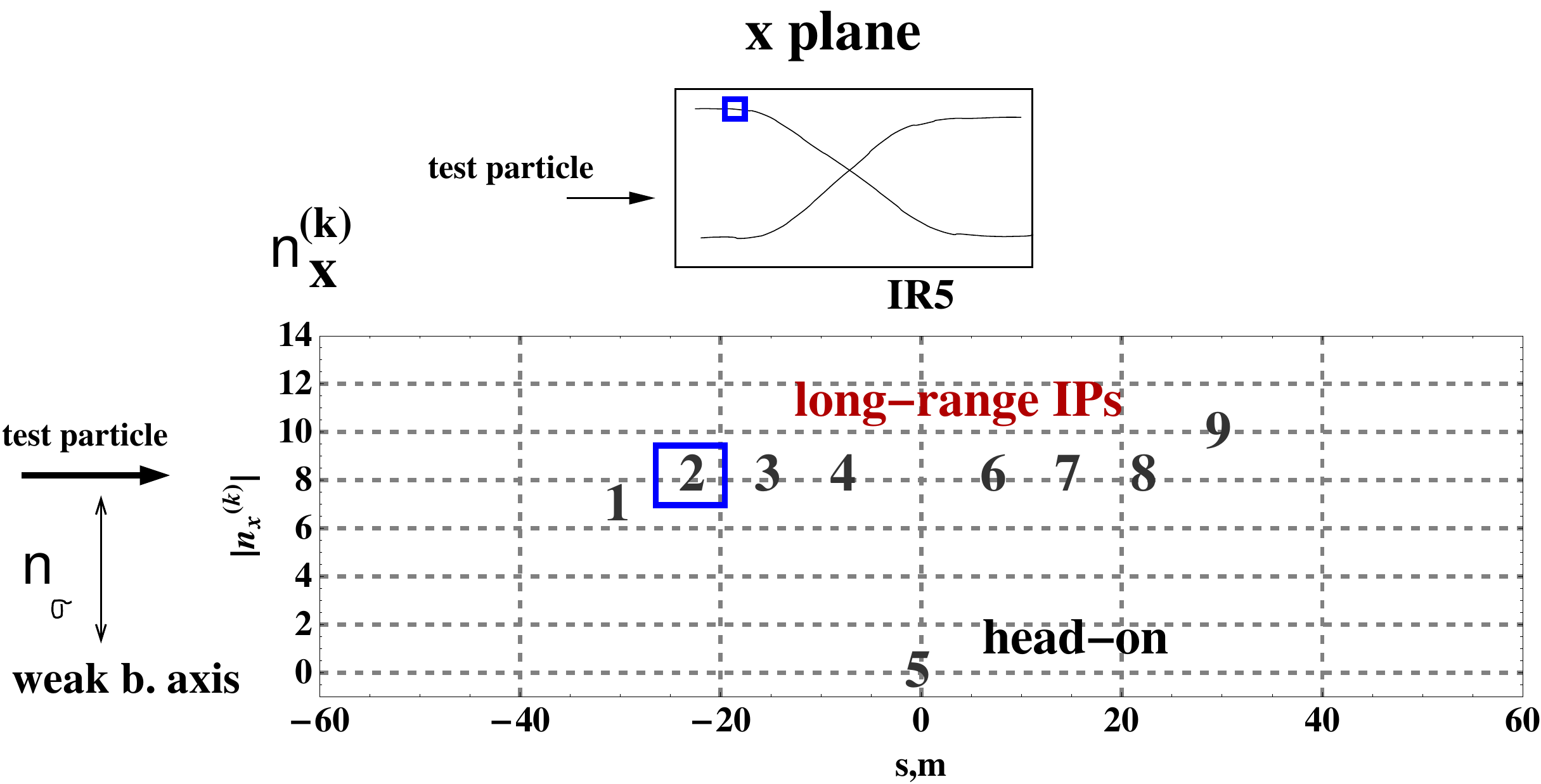}
    \caption{A schematic view of weak- and strong-beam trajectories in
      real (top) and amplitude (bottom) spaces.  A reduced set is
      used: $N_{\lr}=8+8=16$ ($k=1,18$).}
    \label{nxset}
  \end{center}
\end{figure}

The Lie map depends on the above-defined collision set through the
normalized separations $n^{(k)}_{x,y}= d^{(k)}_{x,y}/\sigma^{(k)}$ and
the unperturbed horizontal betatronic phases $\phi^{(k)}$ at the
IPs. Here, $d_{x,y}$ is the real-space offset of the strong-beam
centroid in the $x$ or $y$ direction, and it has been assumed that both the
weak- and strong-beam transverse distributions are round Gaussians of
the same r.m.s.  That is:
\begin{equation}
  \sigma^{(k)} = \sqrt{ \beta^{(k)} \epsilon}  \ \ \ \
  (\beta^{(k)}_x=\beta^{(k)}_y\equiv \beta^{(k)} ) .
  \label{round}
\end{equation}
In Eq.~\re{round}, $\beta^{(k)}$ are the beta functions and $\epsilon$ is
the emittance. It will be shown below that off-plane collisions
contribute very little to smear; thus after excluding these, the
problem becomes one-dimensional and may easily be illustrated
(see Fig.~\ref{nxset}). Here, $n^{(k)}_{x}$ are the strong-beam centroids
in amplitude space: points $(s^{(k)}, n_x^{(k)})$, with $s$ being the
distance to IP5 in metres. 
\subsection{The Beam--Beam Hamiltonian}
For a single collision (see Eq.~\re{oneip}), by omitting the superscript $k$ in
$\sigma$ and $n_{x,y}$, the $x$-motion is described by a kick factor
$F$ (or Hamiltonian $H$) \cite{pac09}:
\begin{eqnarray}
  F=-H(x)&=& \int_0^P (1-\mathrm{e}^{- \alpha})\frac{\mathrm{d}\alpha}{\alpha}\, = \label{hint} \\
  &=& \overline \gamma+ \Gamma_0(P) +\mathrm{ln}(P) \label{hlog} \,,
\end{eqnarray}
\begin{equation}
  P=P(x)=\frac{1}{2}\,\left[
    ( n_x+\frac{x}{\sigma})^2 + n_y^2 \right]\,,\no
  \label{p1}
\end{equation}
where $F$ is in units of $\bb\equiv\frac{N_b r_0}{\gamma}$, $r_0$ is
the classical particle radius, $\Gamma_s(P) \equiv \Gamma(s,P)$
denotes the upper incomplete gamma function \cite{enotes1}, and
$\overline \gamma$ = 0.577216 is Euler's constant. The
corresponding beam--beam kick is as follows:
\begin{eqnarray}
  & \Delta x' \equiv  \frac{\mathrm{d}}{\mathrm{d}x} F(x) = \frac{\partial F}{\partial P}  \frac{\mathrm{d}P}{\mathrm{d}x} = \no \\
  =& \frac{ 2(x+n_x\sigma)  }{ (x+n_x\sigma)^2+  (n_y\sigma)^2 }
  \left[ 1-\mathrm{e}^{- \frac{  (x+n_x\sigma)^2+  (n_y\sigma)^2 }  {2
        \sigma^2}       } \right] .
  \label{kick}
\end{eqnarray}
The Fourier expansion of $H$ is as follows:
\begin{equation}
  H(\en,\phi) =\sum_{m}{\cal C}_m \mathrm{e}^{i m \phi} \,,
  \label{cm}
\end{equation}
where ${\cal C}_m \equiv \frac{1}{2\pi}\int_0^{2\pi} \mathrm{e}^{- i m \phi}
H\; \mathrm{d} \phi$.  These coefficients are easily computed numerically by
using the implementation of $\Gamma$ in {\em Mathematica}
\cite{pac09}. Further, analytical expressions in the form of single
integrals over Bessel functions have been derived in Ref.~\cite{notecoef}.
We display these again in the simplified case $n_y=0$ (no off-plane
collisions):
\begin{eqnarray*}
  & \left.{\cal C}_m \right|_{n_y=0} =
  \displaystyle \int_0^1 \frac{\mathrm{d}t}{t} \times \\
  &
  \begin{cases}
    \displaystyle [ 1-\mathrm{e}^{-\frac{ t }{2}n_x^2} \mathrm{e}^{-\frac{t}{4} \en^2}
    \sum_{k=-\infty}^{\infty} I_{-2k}(t\, \en n_x )
    I_{k}(-\frac{ t}{4} \en^2 ) ]& \\
    \mbox{if } m=0 \mbox{ \ \ and}\\
    \displaystyle -\mathrm{e}^{-\frac{ t }{2}n_x^2} \mathrm{e}^{-\frac{t}{4} \en^2}
    \sum_{k=-\infty}^{\infty} i^{m} I_{m-2k}(t\, \en n_x )
    I_{k}(-\frac{ t}{4} \en^2 ) &\\ \mbox{if } m \neq 0 .
  \end{cases}
\end{eqnarray*}
In the head-on case ($ n_x^{(k)} = n_y^{(k)} =0$), the coefficients
${\cal C}_m$ reduce to the $c^{\rm (ho)}_m$ from Ref.~\cite{chaolect}.
Note that in the most interesting case, amplitudes near the dynamic
aperture, both $\en$ and $n_x$ and hence the Bessel function arguments
are large ($ \gg 1$).

Our first step is to remove the linear and quadratic parts $F_{(1)} =
\left.\frac{\partial F}{\partial x}\right|_{x=0} x $ and $F_{(2)} =
\frac{1}{2} \left.\frac{\partial^2 F}{\partial x^2}\right|_{x=0}\,
x^2$. The non-linear kick factor and the corresponding kick are as follows:
\begin{eqnarray}
  &&  \Fnon= F -F_{(1)} -F_{(2)}\, \label{hnon}\,,  \label{fnon} \\
  && \Delta x'_{\rm nonl} \equiv  \frac{d }{d x} \Fnon(x) \,. \no
\end{eqnarray}
As a next step, we rewrite Eq.~\re{hnon} in action-angle coordinates
$\jx,\phi$ by substituting in it $x= \sqrt{2 \jx \beta} \sin{\phi} =
\en \sigma \sin{\phi} $, where $\en = \sqrt{2 I } = \sqrt{2 \jx/
  \epsilon } $ is the test particle amplitude (Eq. (A.1)).
  Next, we expand in Fourier series:
\begin{eqnarray}
  &\displaystyle \Fnon( \en\, \sigma\,   \sin\phi )=c_0 + \sum_{m\neq 0} c_m \mathrm{e}^{i m \phi}  \,.
  \label{coef1}
\end{eqnarray}
The coefficients $c_m$ are naturally the same as ${\cal C}_m$ above,
with the exception of $c_{1}$ and $c_{2}$, which contain additional
$\mathrm{sin}$ and $\mathrm{sin}^2$ terms (see Eq. (A.1)). 
\subsection{Lie Map and Invariant}
For an arbitrary set of collisions $n_x^{(k)}$, $\phi^{(k)}$ ($k=1,N$),
we represent the LHC lattice by a combination of linear elements and
non-linear kicks. It is shown in the Appendix that, to first order in
$\bb$, the Lie map has the same form as the one for a single kick
\re{oneip} -- where, however, the factor $F$ is given by the sum
$$ F \equiv \sum_{k=1}^{\nip} \Fnon^{(k)}(\en,\phi)$$
and $\Fnon^{(k)}$ are such that, compared to Eq.~\re{coef1}, the $k$th IP
participates with a phase shifted by $\phi^{(k)}$:
\begin{eqnarray}
  &&  \Fnon^{(k)}(\en,\phi)\equiv \left.\Fnon^{(k)}(x)\right|_{x\;
    \rightarrow \en\, \sigma^{(k)}\, sin{(\phi + \phi^{(k)})}} =\no \\
  &&=  \sum_{m\neq 0} C_m^{(k)} \mathrm{e}^{i m \phi} \,.
  \label{ftilde}
\end{eqnarray}
The shift in phase means that the coefficients in Eq.~\re{ftilde} are
simply related to $c_m^{(k)}$: $C_m^{(k)}\equiv c_m^{(k)} \mathrm{e}^{i m
  \phi^{(k)}}$ and still satisfy $C_{-m} = C_m^\star $.  Another
important property of the expansion is that only the oscillating part
is taken (the $m=0$ term is excluded). The invariant for multiple
collision points is as follows (see the Appendix):
\begin{eqnarray*}
  h(I,\phi)= -\mu \jx - \bb \displaystyle \sum_{k=1}^{N}
  \sum_{m=1}^{\infty}  \frac{m\, \mu\, {c}^{(k)}_m (I)}{2 \sin{
      (\frac{m \mu}{2})}} \mathrm{e}^{i m (\phi+\mu/2  + \phi^{(k)})} &\\
  +\;c.c.&
\end{eqnarray*}
The surface of the section in phase space is given by $h(I,\phi)=\mathrm{const}$.
A natural initial condition is now imposed: that the initial point in
phase space for a particle starting at $x_0= \en \sigma$ -- that is, with an
amplitude $I_0 \equiv J_0/\epsilon = \en^2/2$ -- lies on the curve
representing the invariant:
\begin{equation}
  h(\I,\phi) = h(\I_0,\pi/2),
\end{equation}
For a fixed $I_0$, this equation implicitly defines $I$ as a function
of $\phi$.  It satisfies the initial condition $I(0)=I_0$:
\begin{eqnarray*}
  &\displaystyle I(\phi) = I_0 +     \sum_{k=1}^{\nip}
  \left(\mathrm{d}I^{(k)}(\phi)  -\mathrm{d}I^{(k)}(0) \right)\,,
\end{eqnarray*}
\begin{eqnarray}
  &  \mathrm{d}I^{(k)}(\phi)  =\\
  &=\displaystyle \frac{\bb}{ \epsilon}   \sum_{m=1}^{M}
  \left( \frac{m\,  { c}^{(k)}_m (I_0)}{2\sin{(m \mu/2)}} \mathrm{e}^{i m (\mu/2
      +\phi - \phi^{(k)} +\pi/2)} + c.c. \right). \no
  \label{dinv}
\end{eqnarray}
Note that, to first order, the argument in ${c}^{(k)}_m$ has been
replaced with $I_0$. We have also separated the two sums so that
$\mathrm{d}I^{(k)}(\phi) -\mathrm{d}I^{(k)}(0)$ is the individual contribution of the
$k$th IP.  In the same way, a different initial condition may be used
(more suitable for plots): $I(0)=I_0$, instead of $I(\pi/2)=I_0$.

The smear $S(\en)$ is now defined as the normalized r.m.s.\@ of the
invariant -- that is, $\sqrt{V}$, with $V$ being the variance:
\begin{eqnarray*}
  &  S(\en)= {\sqrt{V}}/{\langle\,I\,\rangle}\,,  \\
  &    V= \frac{1}{2\pi} \int (I  - \langle I \rangle )^2 \mathrm{d}\phi\,, \ \ \ \
  \langle I \rangle= \frac{1}{2\pi} \int I \mathrm{d}\phi\,.
\end{eqnarray*}
\section{Verification with tracking}
\begin{figure}[h]
  \begin{center}
    \includegraphics[width=0.2\textwidth,angle=0] {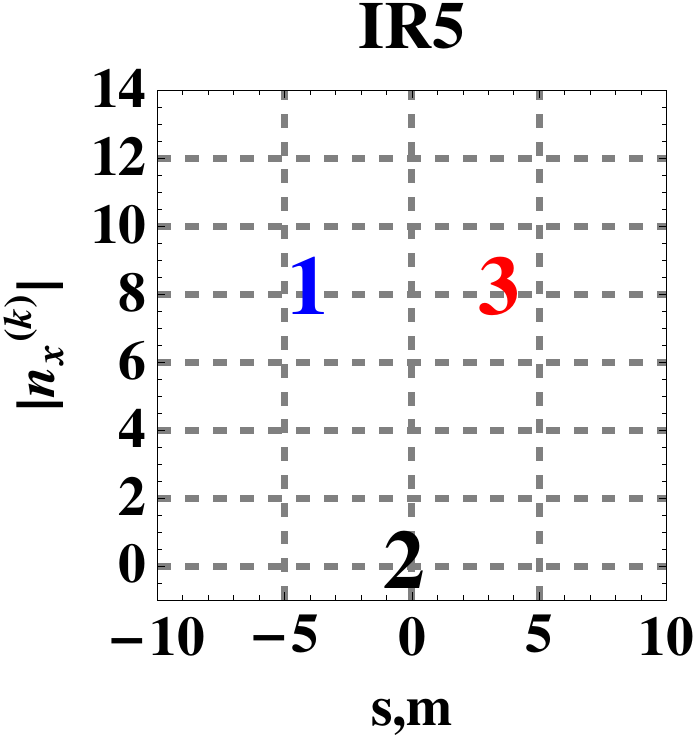}
    \includegraphics[width=0.2\textwidth,angle=0] {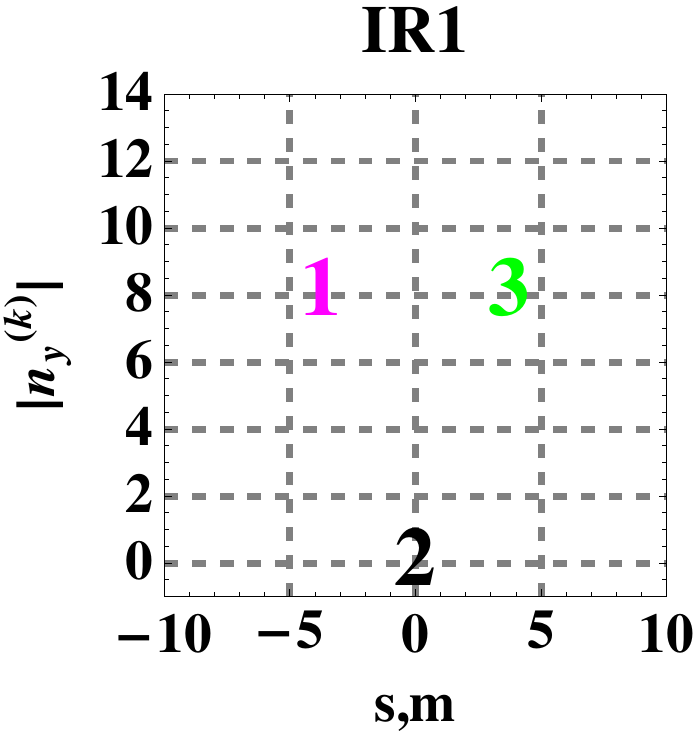}
    \vspace{-.3cm}
    \caption{The sample set-up: three collisions in each IR5 and IR1. The
      l.r. are set at 8 sigma.}
    \label{reduced}
  \end{center}
\end{figure}

\begin{figure}[h]
  \begin{center}
    \includegraphics[angle=0,width=0.33\textwidth]{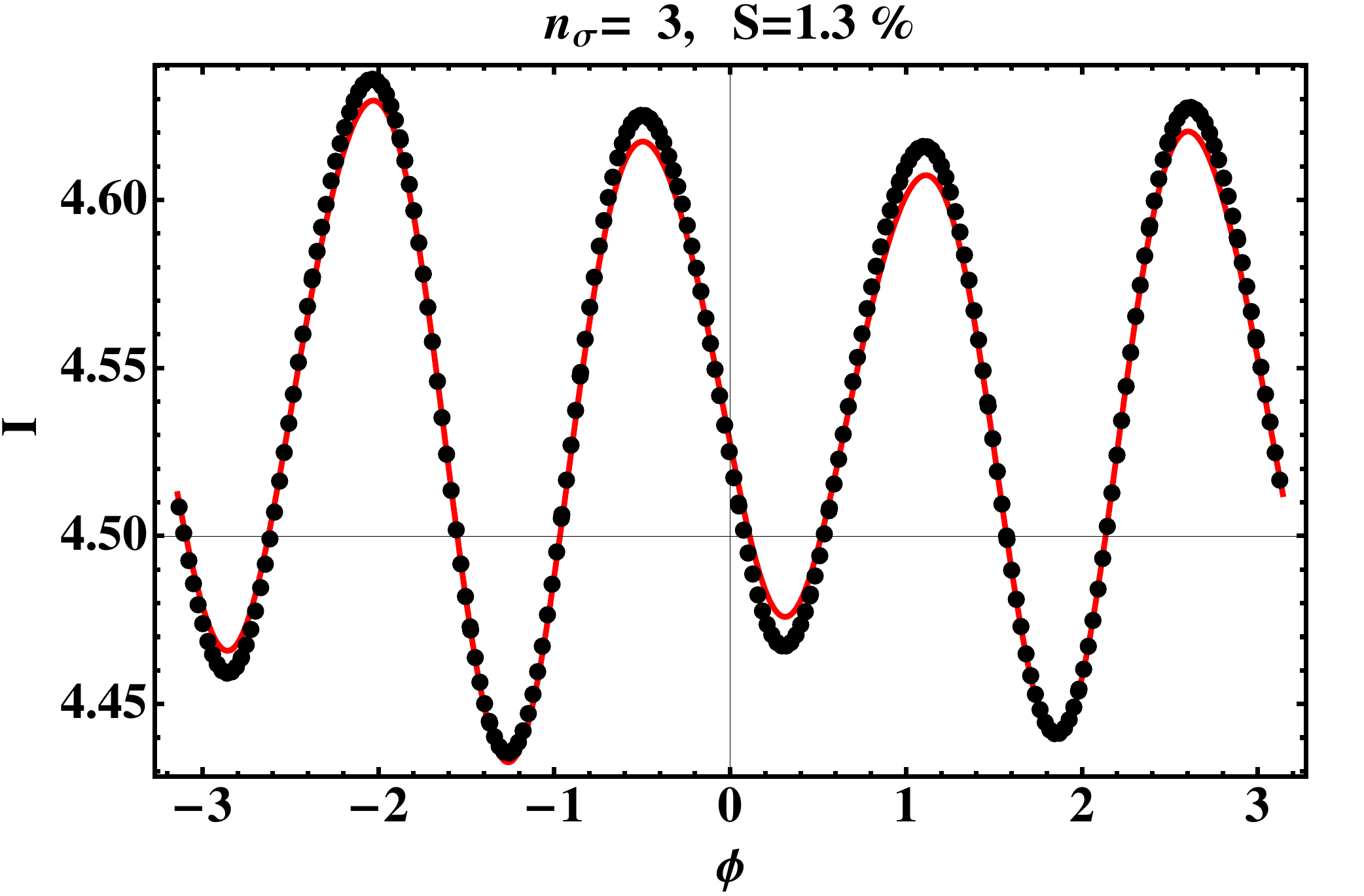}
    \includegraphics[angle=0,width=0.33\textwidth]{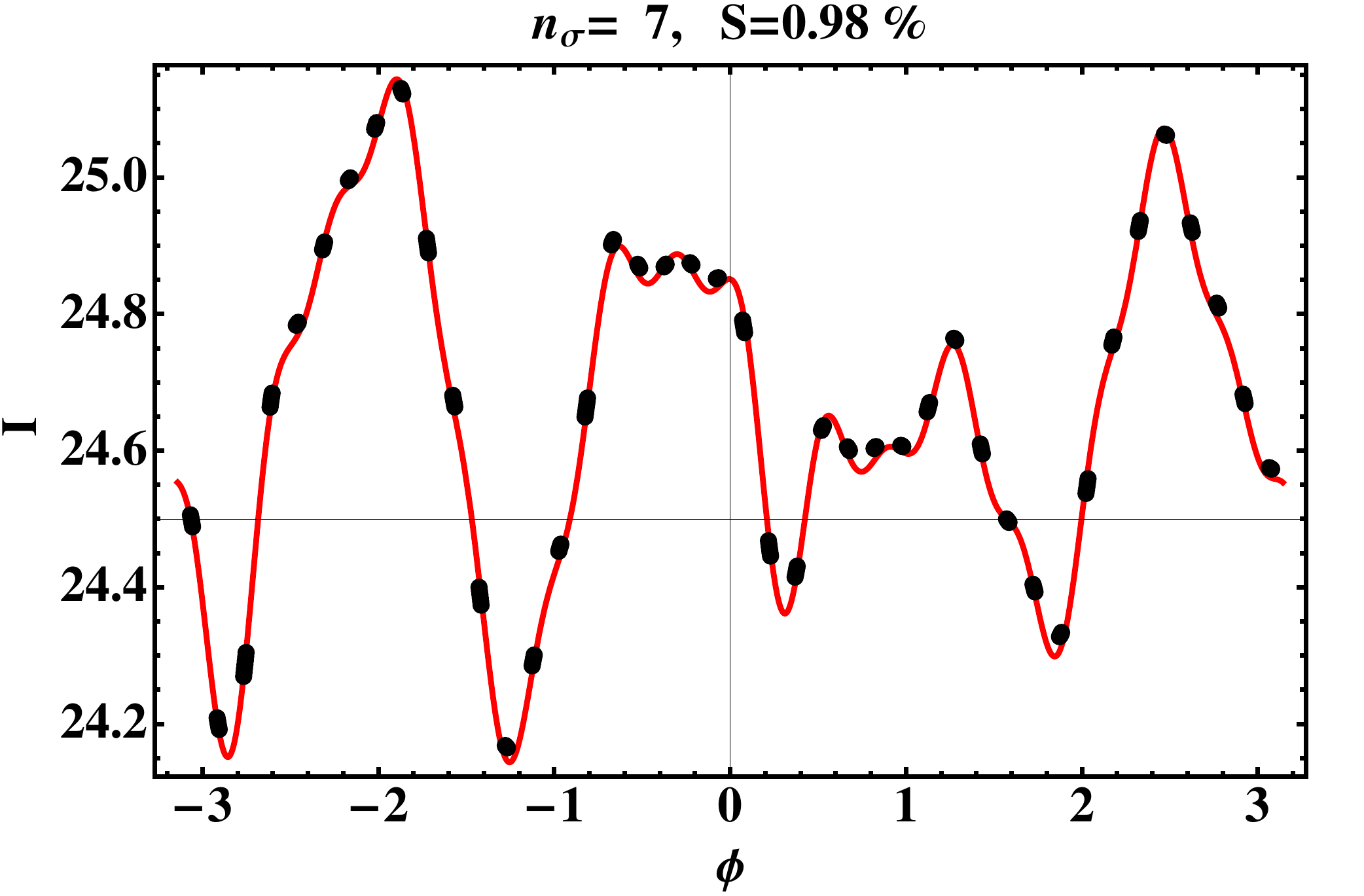}
  \end{center}
  \vspace{-.8cm}
  \caption{An invariant tested on a simple kick-matrix model. Black
    points: turn-by-turn coordinates ($\phi,\I$) for $10^{3}$ turns.
    Red: invariant $I(\phi) $ (initials chosen so that
    $I(\pi/2)=I_0$).} 
  \label{I-7}
\end{figure}

As an example application, this section studies the very simple
collision set that still possesses all the symmetries with the l.r. set at 8
sigma, as depicted in Fig.~\ref{reduced}. Both IR5 and IR1 are
included. The goal here is to test the invariant $I(\phi)$ by tracking
with a simple model built with kicks $\Delta x'_{\rm nonl}$
alternating with linear matrices and SixTrack. The parameters are as follows:
energy $3.5~$TeV, $\np=1.2\times10^{11}$, and normalized emittance
$\epsilon_n=2.5\times10^{-6}$.
Tracking single particles at various amplitudes with the simple model
produces the results shown in Fig.~\ref{I-7}.  A particle starts
with $n_{\sigma}=3$, or $7$ ($I_{0}=4.5$, or $24.5$). The $c_m$ are
computed with an accuracy of $10^{-7}$ -- the value of $M$ in Eq.~\re{dinv} is
about $40$.
Since the beam--beam potential changes the linear optics, we need to
find the linearly perturbed matched $\beta$-function value at the
initial point for tracking.  For the plots in Fig.~\ref{I-7}, this is
done in a separate run, using a linear kick $(\Delta x')_{\rm lin}$
(only terms $\sim x^2$ in the Hamiltonian). This is similar to what is
done in SixTrack. The resultant matched $\beta$ is used to define the
initial coordinate $x_0$ (through $n_{\sigma}$).  The values of the
smear are shown at the top of each plot.

\begin{figure}[h]
  \begin{center}
    \includegraphics[angle=0,width=0.40\textwidth]{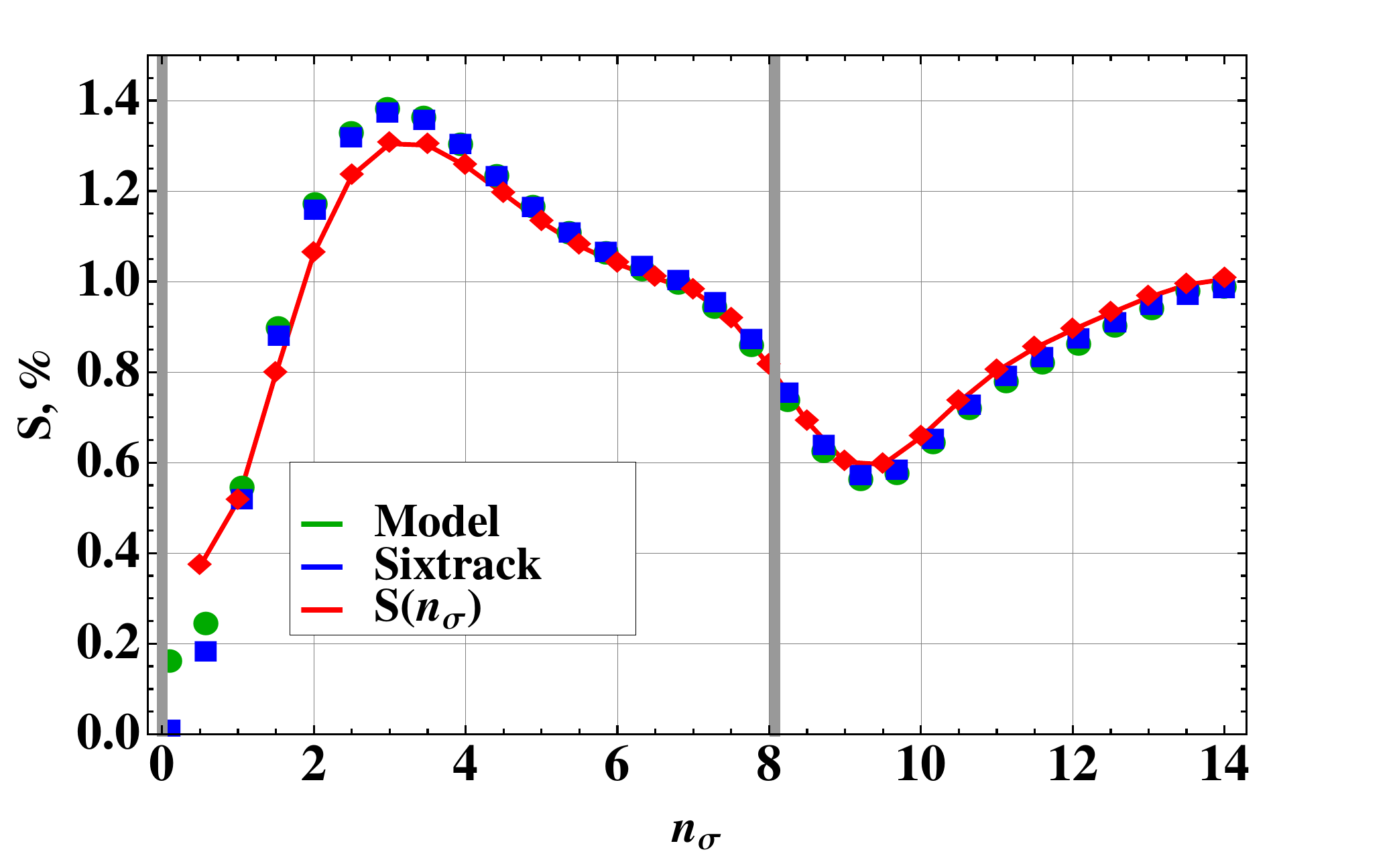}
  \end{center}
  \vspace{-.8cm}
  \caption{Agreement with SixTrack.  }
  \label{figsm}
\end{figure}

Plotting the smear over a range of amplitudes with all three methods --
model, SixTrack, and analytical $S(\en)$ -- results in
Fig.~\ref{figsm}. Note that here the images of the strong-beam
centroids (see Fig.~\ref{nxset}) are represented by vertical grey
lines drawn at $0$ and $8$ sigma.
Let us now look at the individual contributions to $I(\phi)$ of the
six IPs at three amplitudes chosen arbitrarily; say, $\en=1,3$, and $7$.
\begin{figure}[h]
  \begin{center}
    \subfigure[]{\label{5a}\includegraphics[trim=.1cm .1cm .5cm .1cm,
      clip=true, width=0.23\textwidth, angle=0]{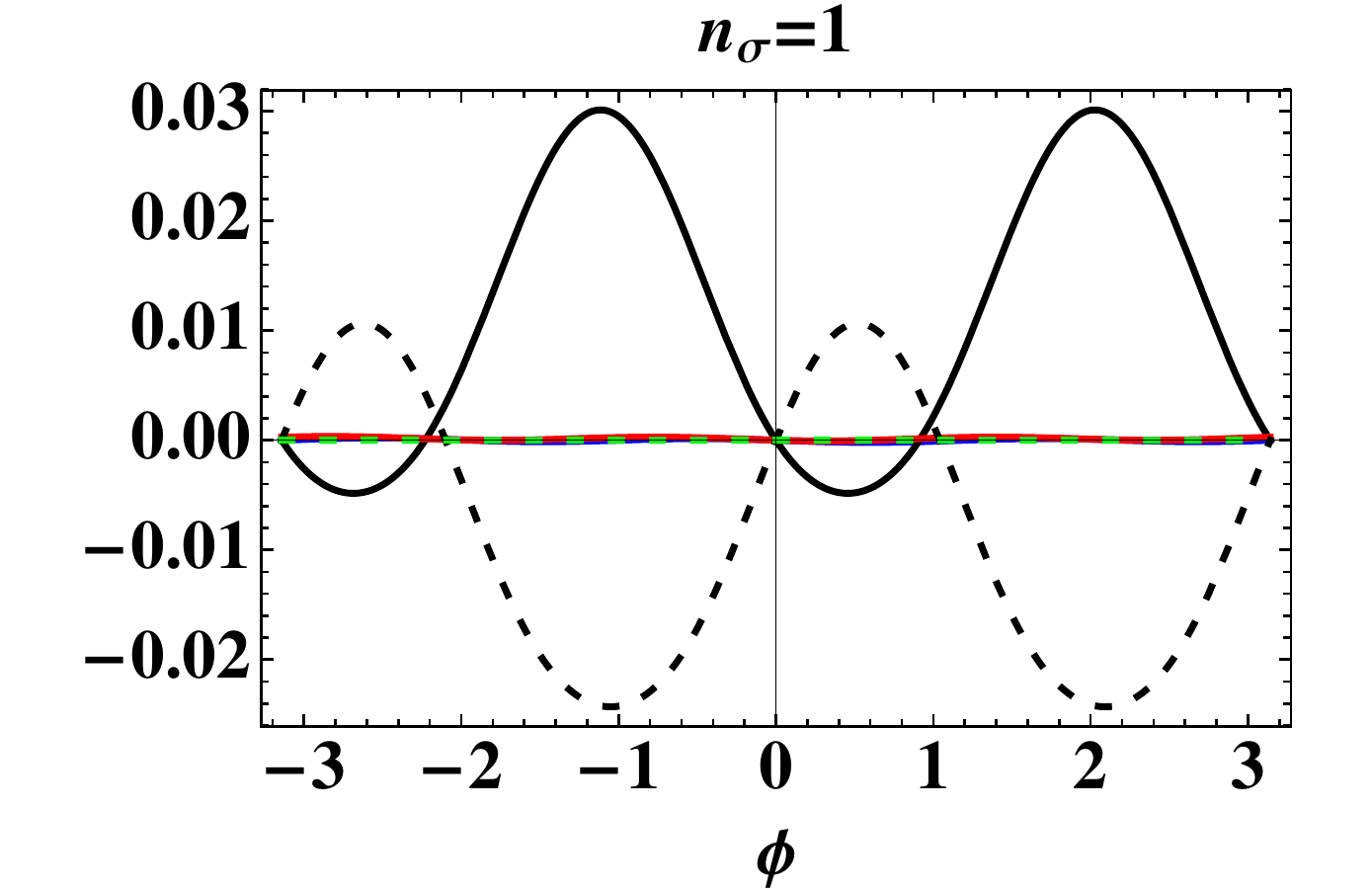}}
    \subfigure[]{\label{5b} \includegraphics[trim=.1cm .1cm .5cm .1cm,
      clip=true, width=0.23\textwidth, angle=0]{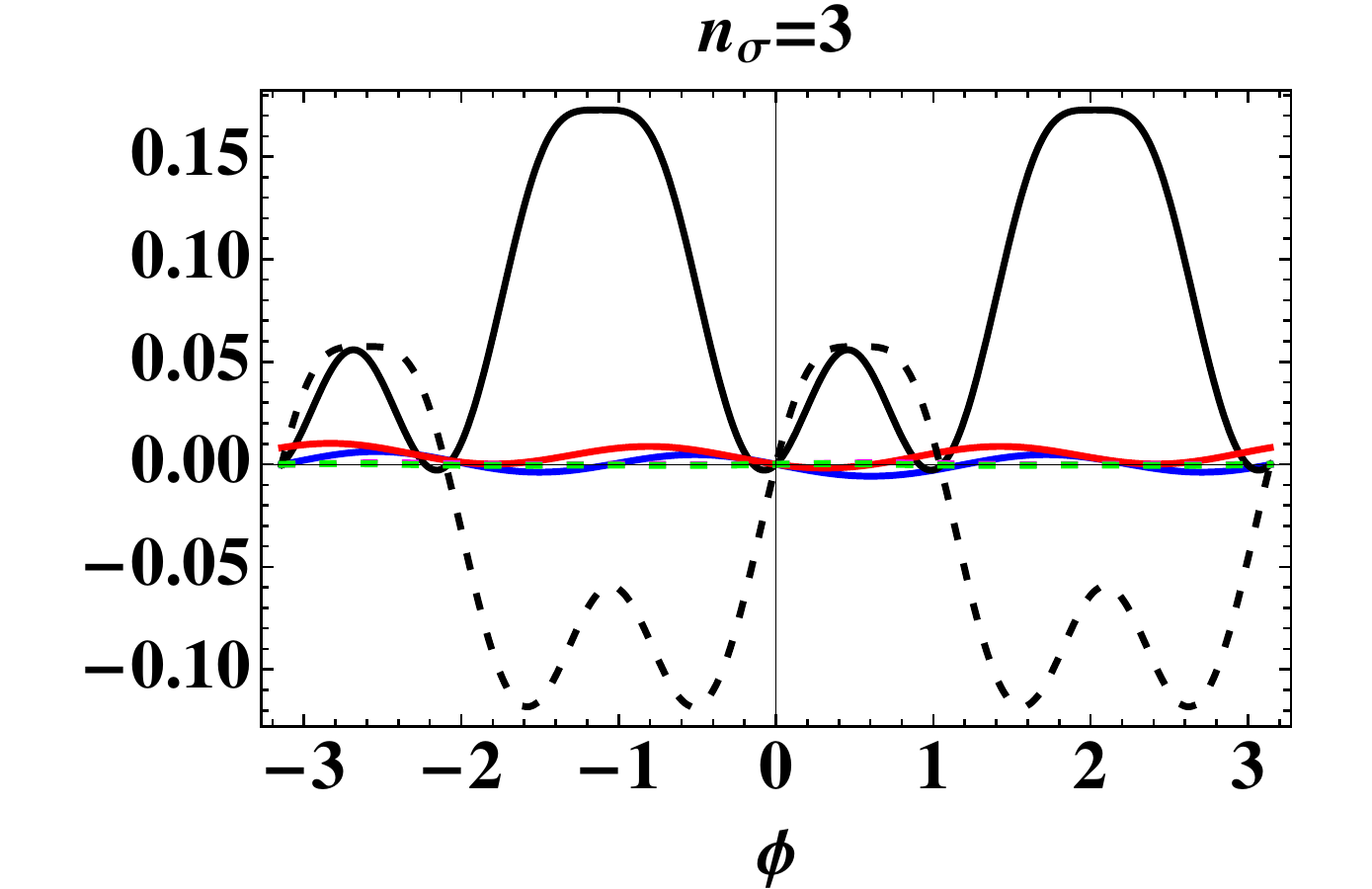}}
    \subfigure[]{\label{5c} \includegraphics[trim=.1cm .1cm .5cm .1cm,
      clip=true, width=0.24\textwidth, angle=0]{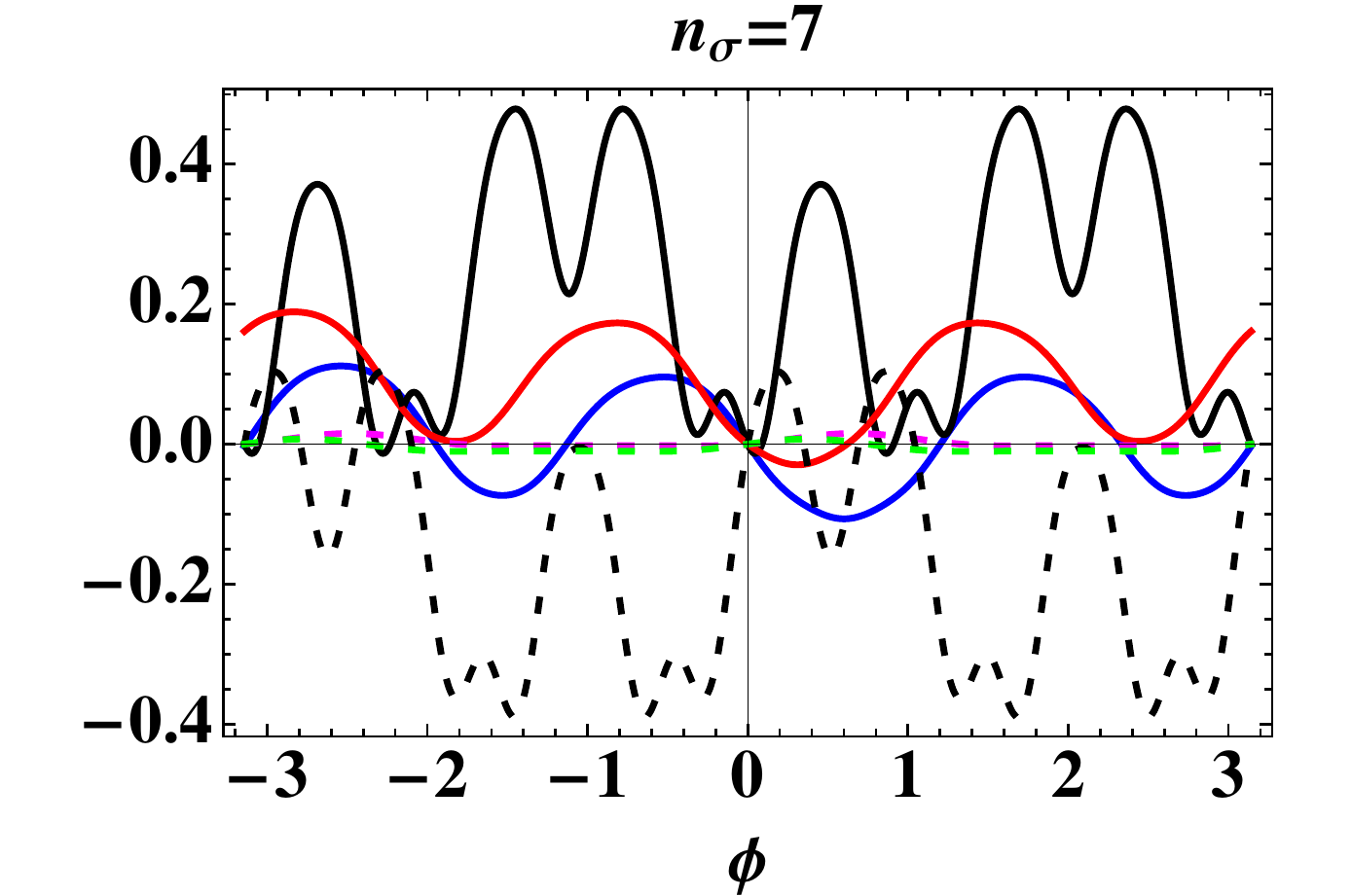}}
    \vspace{-.3cm}
    \caption{Individual contributions $dI^{(k)}(\phi) -dI^{(k)}(0)$ --
      color code as in Fig.~\ref{reduced}. }
    \label{pltc13}
  \end{center}
\end{figure}
The excursions (w.r.t. $I_0$) of the individual invariant surfaces
are shown in Fig.~\ref{pltc13}. Here, $I(0)=I_0$.  The colour code is as
in Fig.~\ref{reduced}, and in addition for the head-ons we use solid
black for IP5 and dashed for IP1.  Near the axis ($\en=1$), only the
two head-ons contribute and, being of opposite signs, almost
compensate each other. At $\en=3$, one begins to see long-range
contributions that grow when $\en=7$. At such large amplitudes, the
compensation is no longer true.  Magenta and green are barely seen,
meaning that the contribution of off-plane collisions is negligible.  Thus
in the case of a test particle moving in the horizontal motion, the
contribution of all l.r. in IR1 can be neglected, and vice versa for
vertical motion and IP5.
\section{The behaviour of the smear $S(\en)$ near the dynamic
  aperture}
Above some critical strength of beam--beam interaction -- that is,
quantities $N_{\lr}$ and/or $\np$ and/or an inverse crossing angle -- the
first-order theory is no longer an adequate description of the smear.
However, as we will see, the behaviour of $S(\en)$ may still be used as
an indication of the dynamic aperture, since it exhibits a local maximum
near it.
\begin{figure}[h]
  \begin{center}
    \subfigure[]{\label{6a} \includegraphics[trim=.4cm .1cm 1.5cm
      1.cm, clip=true, width=0.23\textwidth, angle=0]{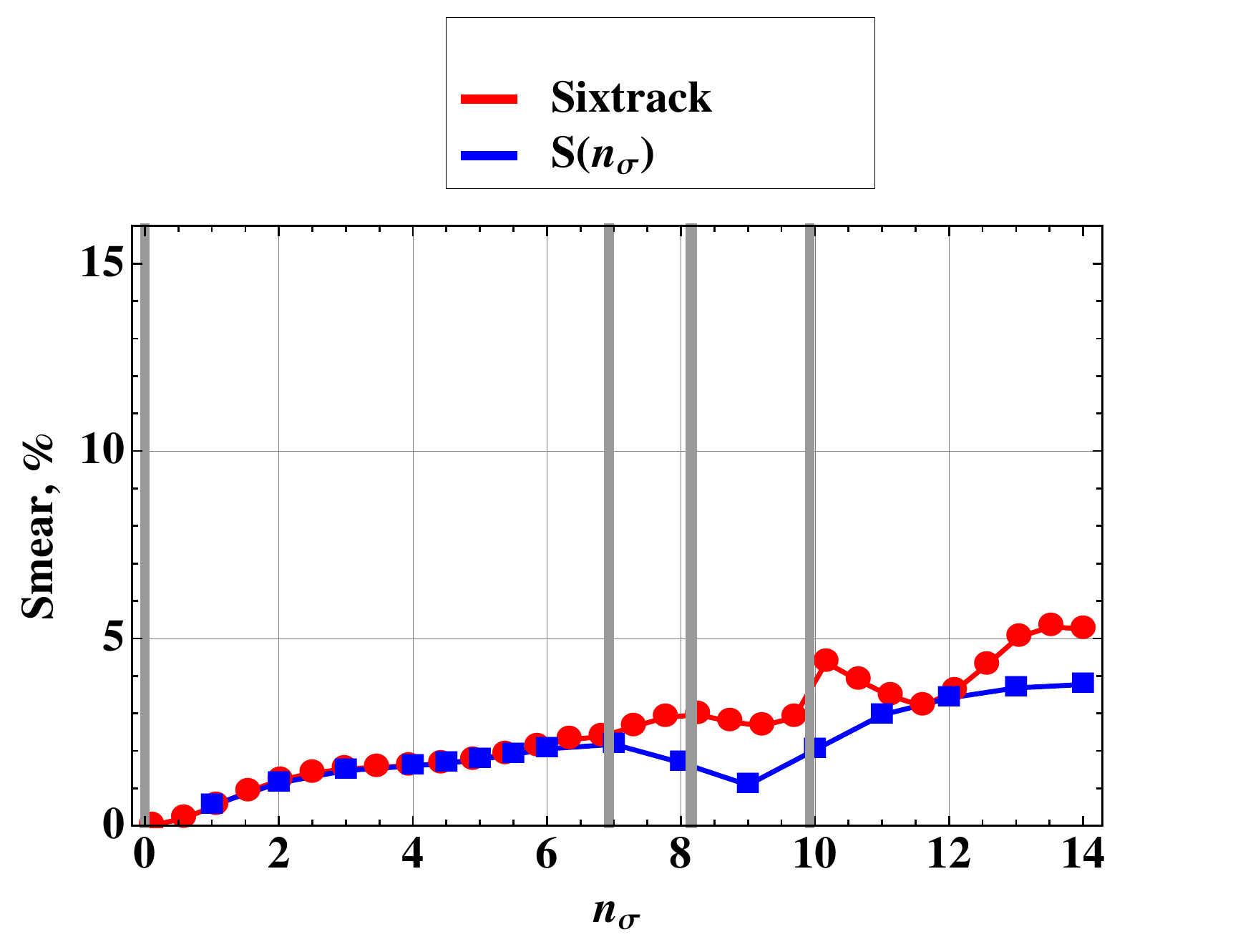}}
    \subfigure[]{\label{6b} \includegraphics[trim=.4cm .1cm 1.5cm
      1.cm, clip=true, width=0.23\textwidth, angle=0]{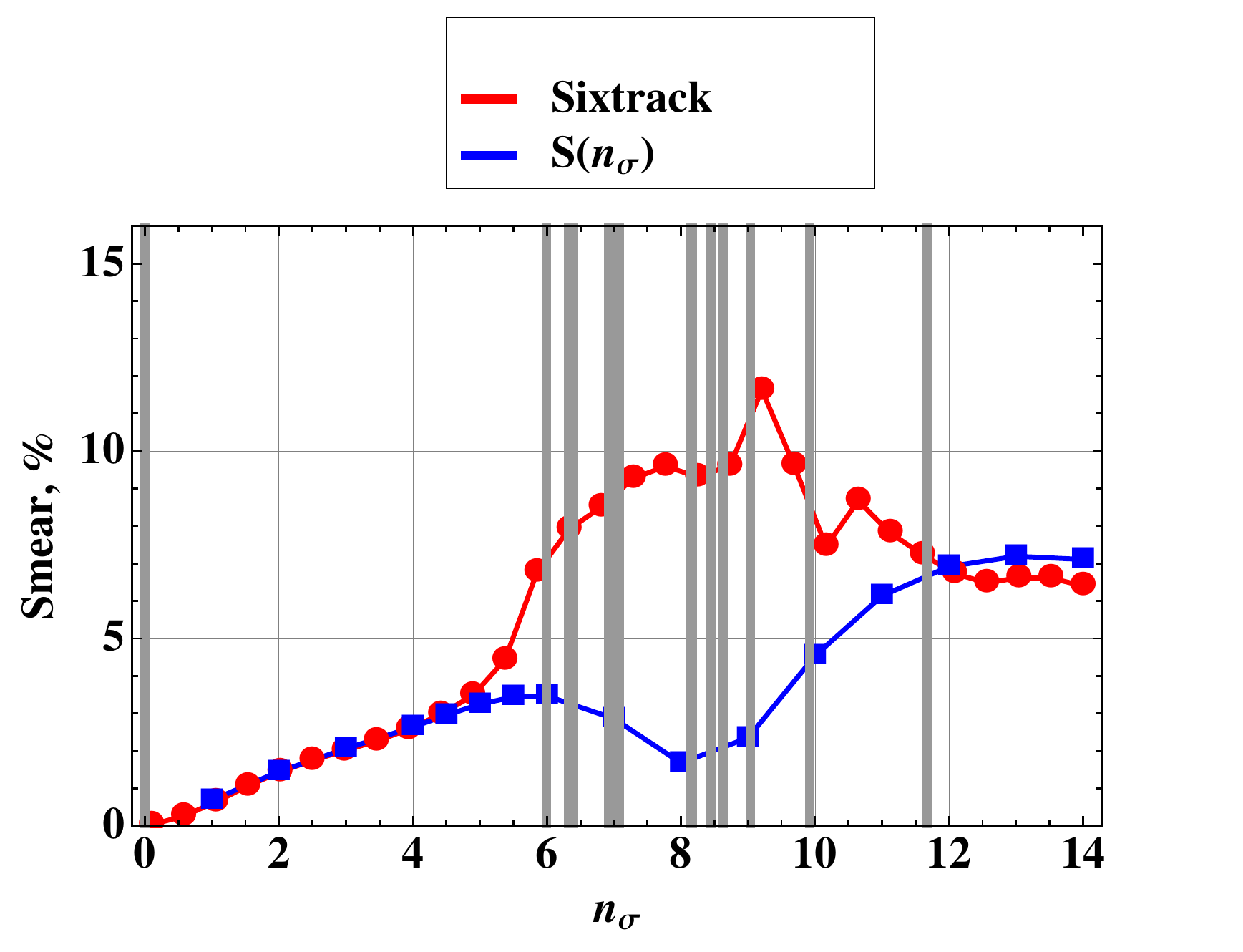}}
    \subfigure[]{\label{6c} \includegraphics[trim=.4cm .1cm 1.5cm
      2.8cm, clip=true, width=0.23\textwidth, angle=0]{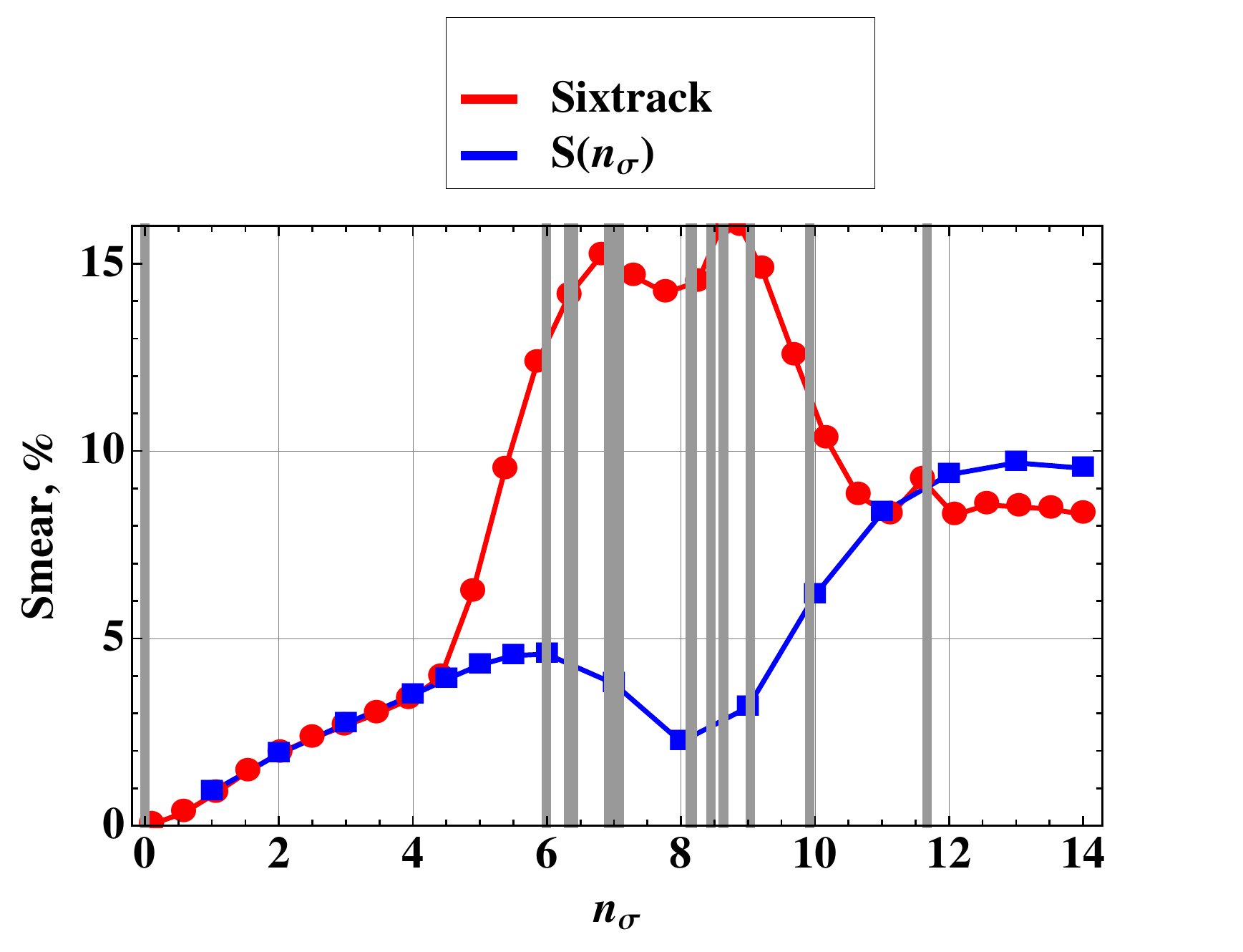}}
    \subfigure[]{\label{6d} \includegraphics[trim=.4cm .1cm 1.5cm
      2.8cm, clip=true, width=0.23\textwidth, angle=0]{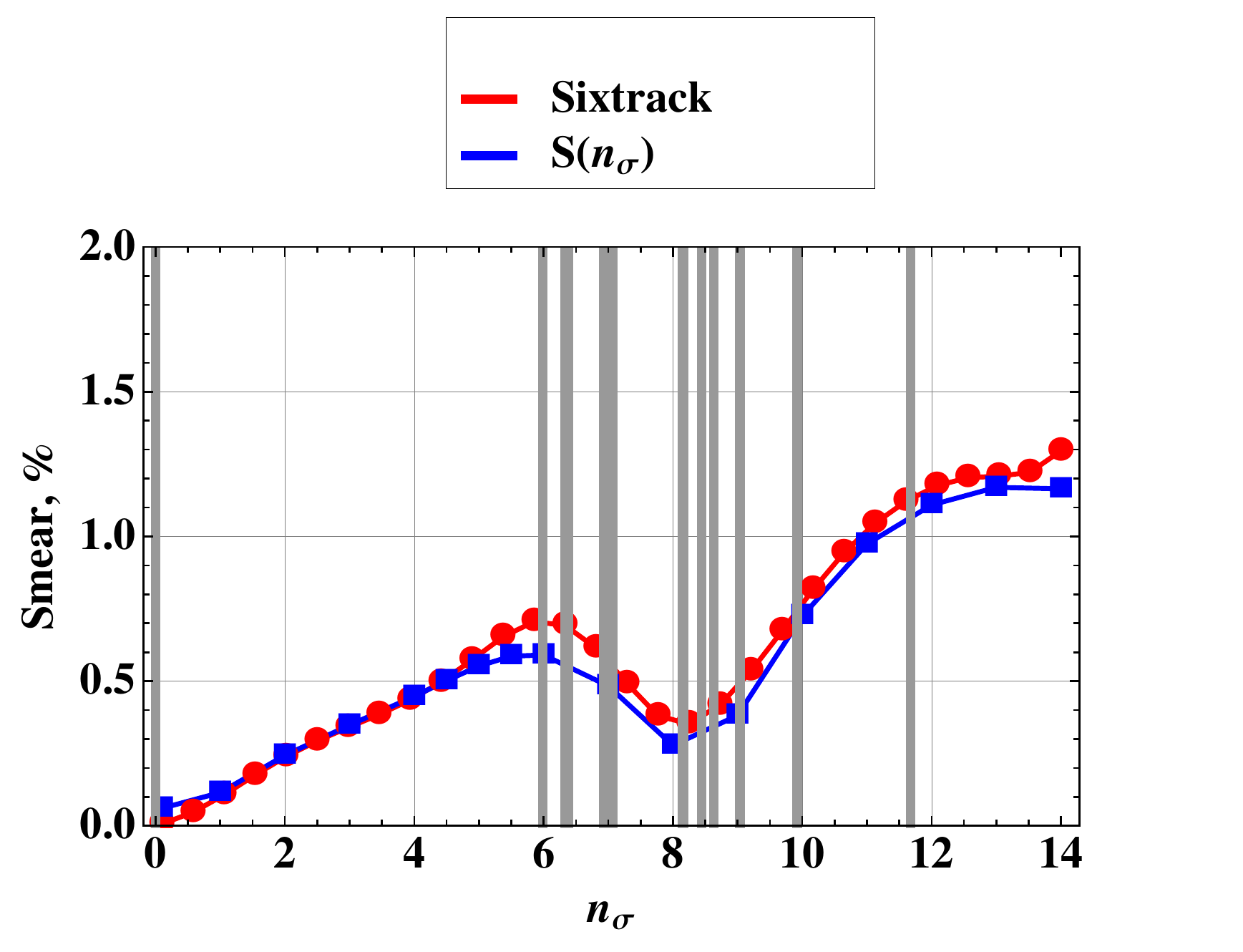}}
    \vspace{-.5cm}
    \caption{Various combinations of numbers of long-range collisions
and bunch intensities to illustrate linear and non-linear behaviour: }
{\footnotesize
    \begin{center}
      \begin{tabular}{c c c }
      \toprule
        & $N_{\lr}$ & $\np$  \\
      \toprule
        (a)& $16$ & $1.2\times10^{11}$\\
      \midrule
        (b)& $32$ & $1.2\times10^{11}$\\
      \midrule
        (c)& $32$ & $1.6\times10^{11}$\\
      \midrule
        (d)& $32$ & $0.2\times10^{11}$ \\
      \bottomrule
      \end{tabular}
    \end{center}
  }\vspace{-1cm}
  \label{sets12}
\end{center}
\end{figure}
What happens is that the linear behaviour -- that is, the agreement between the
first-order $S$ and SixTrack at all amplitudes seen in
Fig.~\ref{figsm} -- is replaced by what is shown in Figs.~
\ref{sets12}(a)--(c).  The blue ($S(\en)$) and the red (SixTrack)
curves depart from each other once $\en$ approaches amplitudes near
the strong-beam core, represented by the cluster of vertical grey
lines. At this point, the exact smear (red) exhibits a steep growth;
thus the dynamic aperture is likely to be close to this point, while
$S$ goes through a maximum and then through a minimum, thus forming a
dip.  Upon exiting the core, past the last grey line, the red and blue
curves almost re-merge.  It can be shown that the above property of
$S(\en)$ is a consequence of the left--right symmetry of IR5 and
IR1. Namely, the individual contributions (such as the red and blue curves
in Fig.~\ref{reduced}) change sign or flip about the axis each time
$\en$ crosses a grey line. At this amplitude, $S(\en)$ stops growing
and goes through a maximum.

\section{Analysis of long-range experiments }
\subsection{Dependence on Intensity and Crossing Angle}
We set the parameters as at the MD: energy $3.5~$TeV,
$\epsilon_n=2.5\times10^{-6}$ \cite{footnote1}, and
$\beta^{\star} = 0.6$~m.

Of all the collision sets used at the MD, let us consider three:
$N_{\lr}=32$, 24, and 16.  For each of them, two parameters, the bunch
intensity $\np$ and the (half) crossing angle $\alpha$, uniquely define the
dependence of the first-order smear on amplitude
$S(n_{\sigma};\np,\alpha) $ through the following procedure.  First,
being a first-order quantity in $\bb$, the smear is obviously
proportional to the intensity: $S \sim \np$.  Second, the dependence
of $n_{x,y}^{(k)}$ on the (half) crossing angle $\alpha$ is given by
the well-known scaling law: $n_{x,y}^{(k)}\sim \alpha
\sqrt{\beta^{\star}}$, where $n_{x,y}^{(k)}$ are taken from some
sample lattice built for $\beta^{\star} = 0.55$~m and $\alpha=125$.
Finally, the phases $\phi^{(k)}$ are assumed to be independent of
$\alpha$.

\begin{figure}[ht]
  \begin{center}
    \includegraphics[width=.45\textwidth,angle=0]{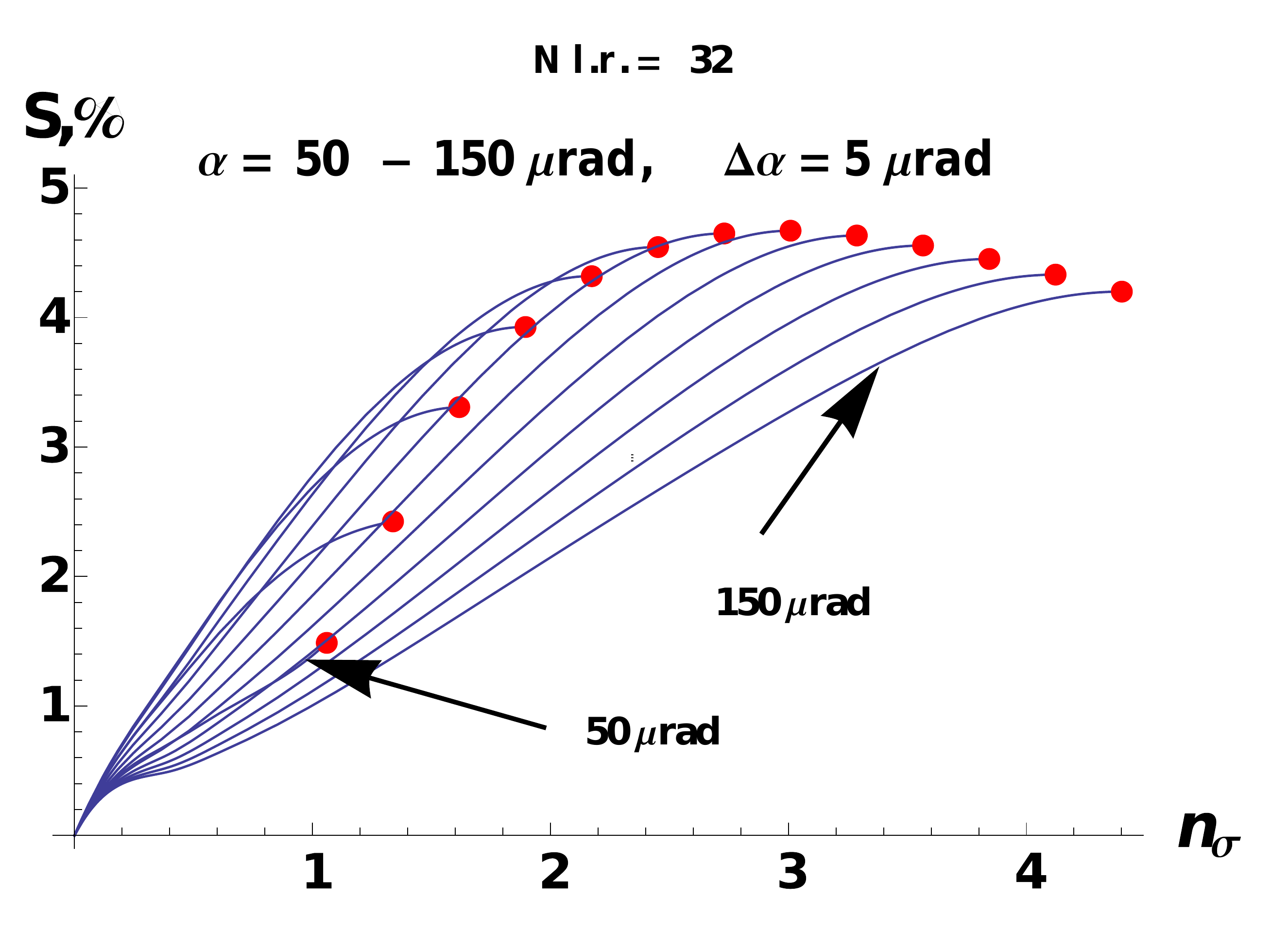}\\
    \includegraphics[width=.24\textwidth,angle=0]{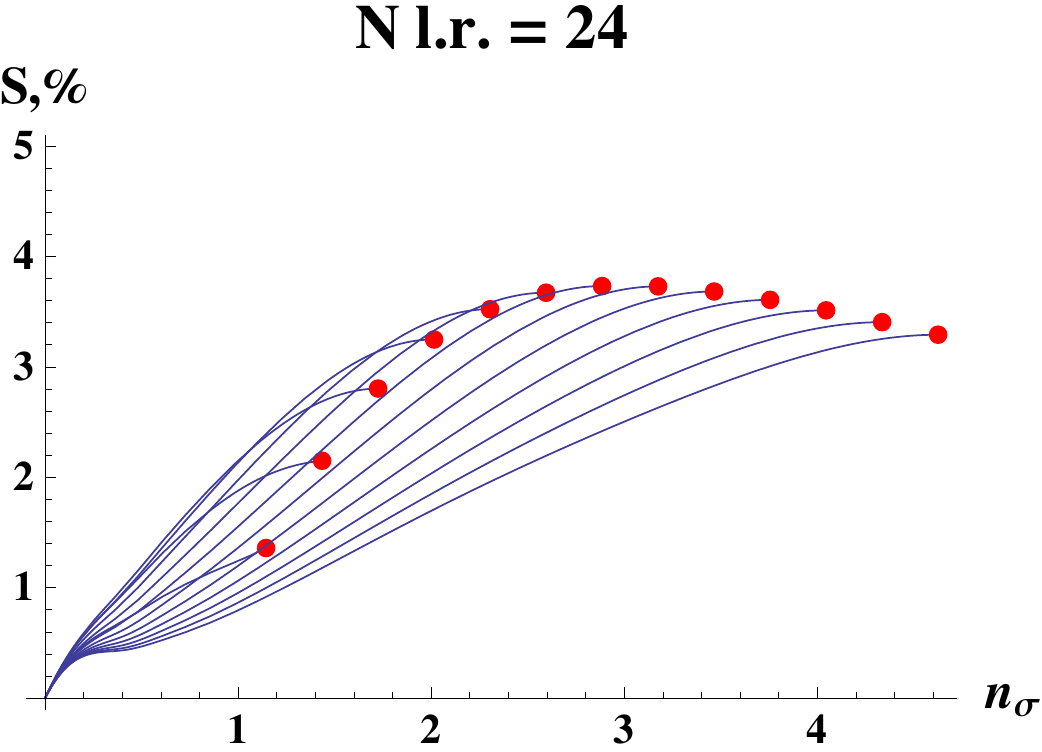}
    \includegraphics[width=.24\textwidth,angle=0]{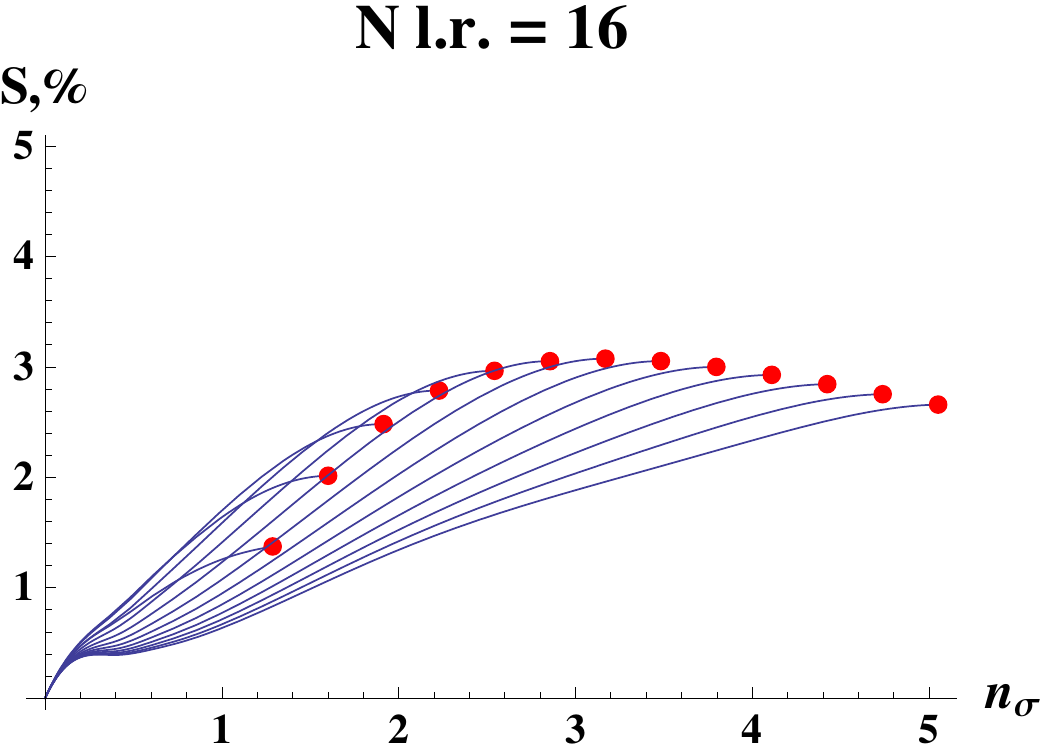}
  \end{center}
  \vspace{-.5cm}
  \caption{The dependence of the smear(amplitude) graph on the parameter
    $\alpha$ for $N_{\lr}=32$ (top) and $N_{\lr}=24$, $16$
    (bottom). Each graph is restricted within a domain extending up to
    its first maximum (red dot) (the entrance into the strong-beam core).}
  \label{deponang}
\end{figure}

The dependence on the angle is presented in Fig.~\ref{deponang}.  Each
blue branch corresponds to $S(n_{\sigma};1.2\times10^{11}, \alpha)$
being taken over an amplitude range where it is monotonically increasing;
hence, as we already know, it will remain in agreement with the tracking
for any strength of the beam--beam interaction.
\begin{figure}[h]
  \begin{center}
    \includegraphics[width=0.45\textwidth]{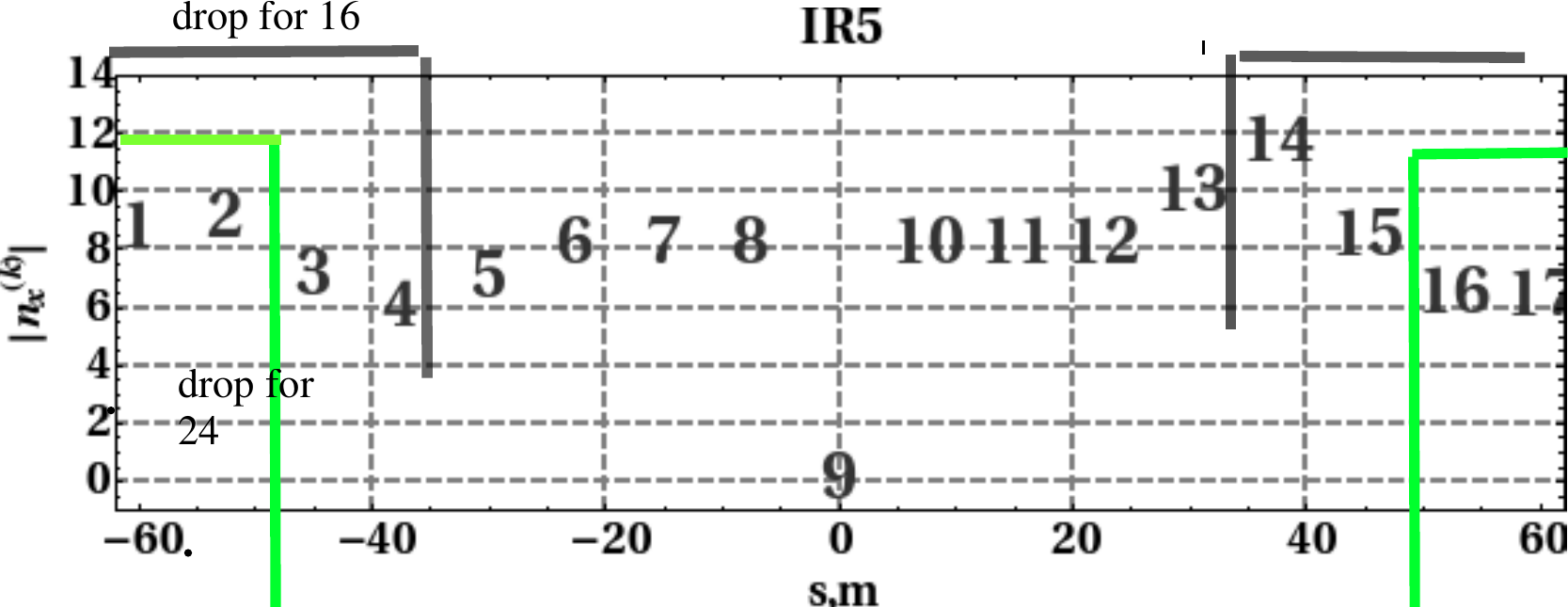}
    \caption{The collision sets for $N_{\lr}$=24 and 16 are built by
      dropping the first and last two or four elements from the full set
      ($N_{\lr}=32$).}
    \label{drops}
  \end{center}
\end{figure}

\begin{figure}[h]
  \begin{center}
    \includegraphics[width=0.5\textwidth]{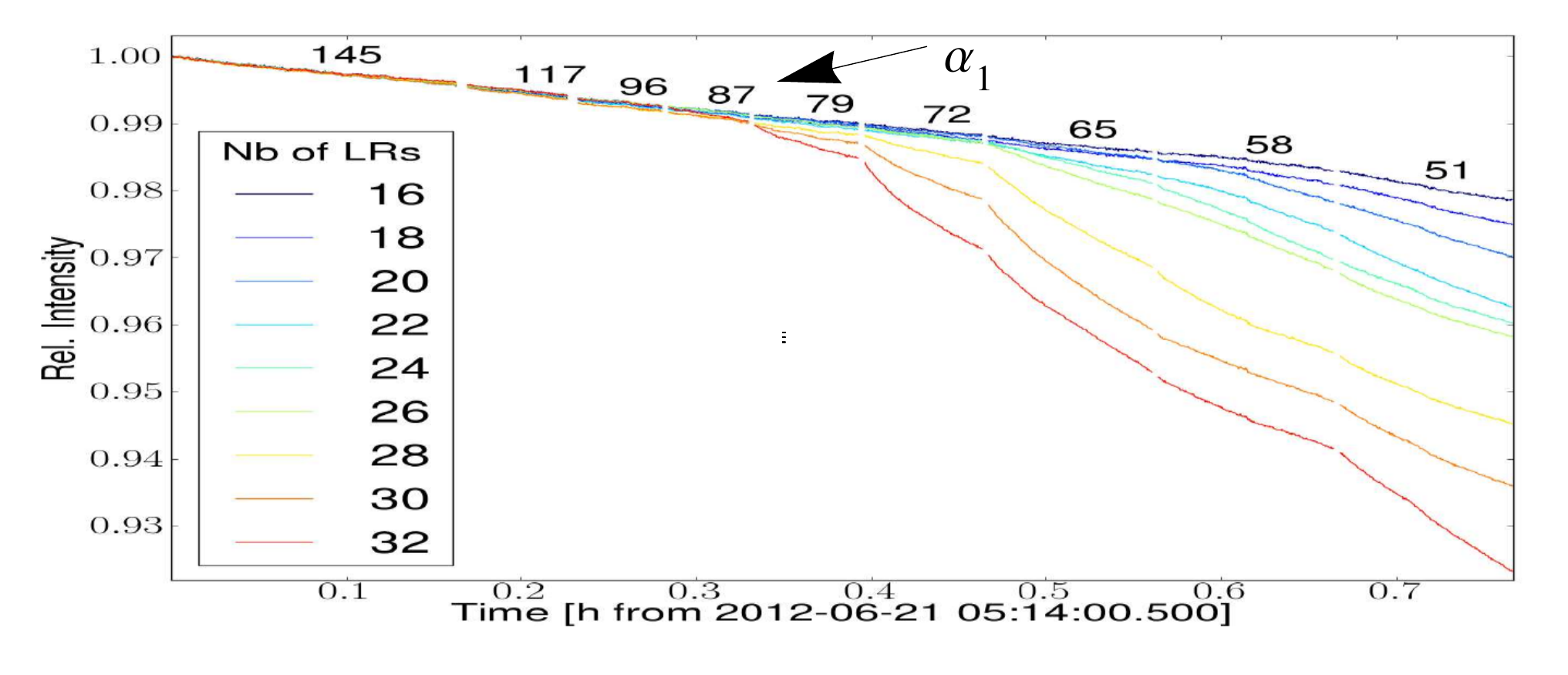}
    \caption{An experiment with $\np=1.2\times10^{11}$: losses start at $
      \alpha_1 \approx 87 \; \murad$.}
    \label{exp2}
    \includegraphics[width=0.5\textwidth]{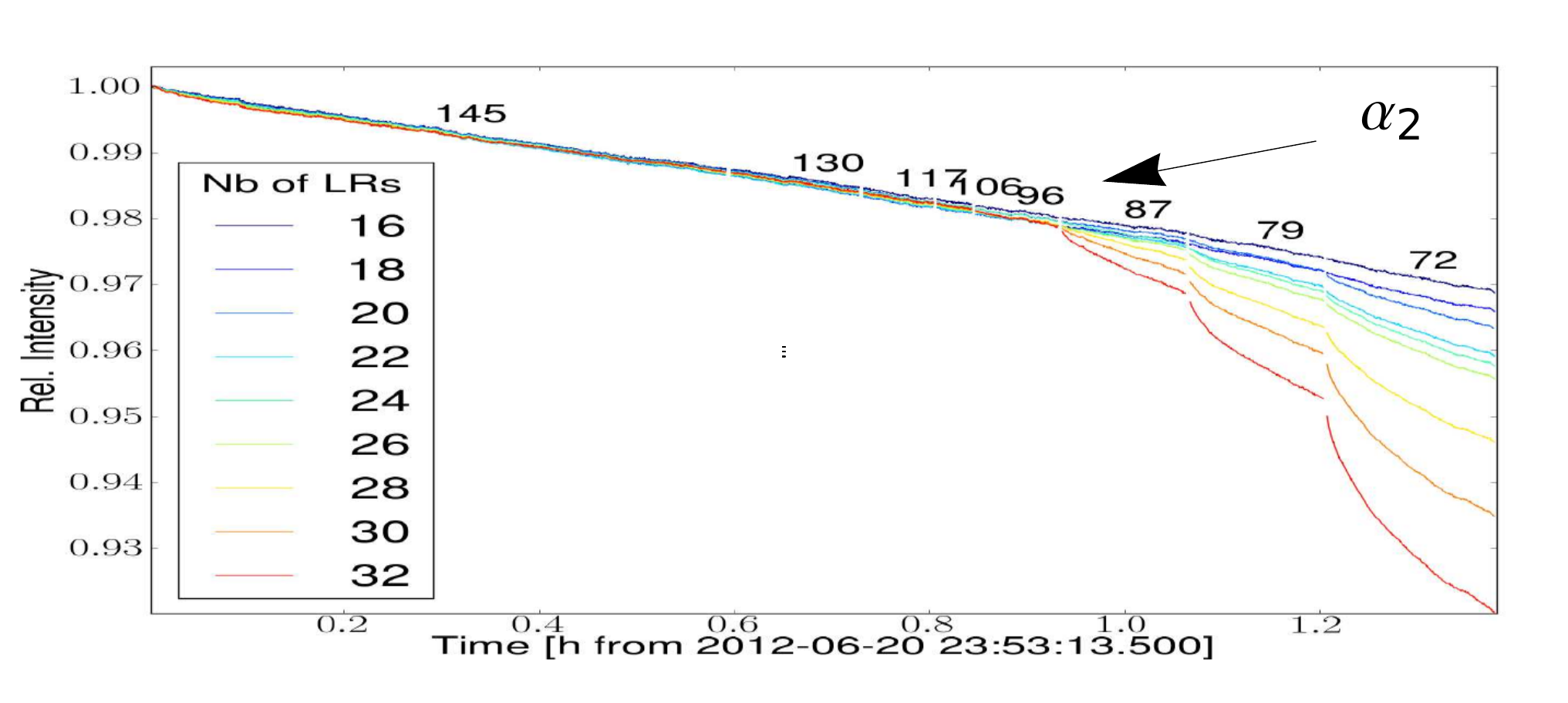}
    \caption{An experiment with $\np=1.6\times10^{11}$: losses start at
      $\alpha_1 \approx 96 \murad$.}
    \label{exp1}
  \end{center}
\end{figure}

Coming now to the MD, the observed losses during reduction of the
crossing angle in IP1 are shown in Figs.~\ref{exp2} and \ref{exp1}
\cite{mdnote}.
\subsection{An Explanation of the Case $N_{\lr}=32$ (Brown Curves)}
For $N_{\lr}=32$ (the full $50$-ns collision set shown in
Fig.~\ref{drops}), we need to explain the brown curves in
Figs.~\ref{exp2} and \ref{exp1}. Here, losses are seen to start at
$\alpha_1\approx 87$ and $\alpha_2\approx 96 \murad$, respectively.

\begin{figure}[H]
  \includegraphics[width=.24\textwidth,angle=0]{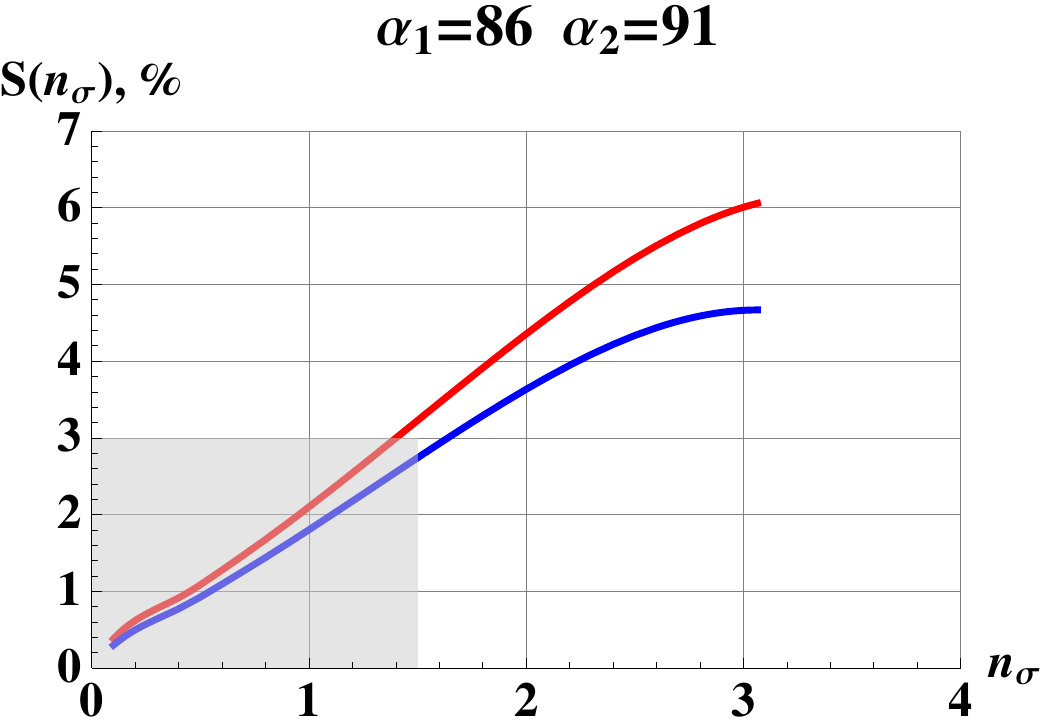}
  \includegraphics[width=.24\textwidth,angle=0]{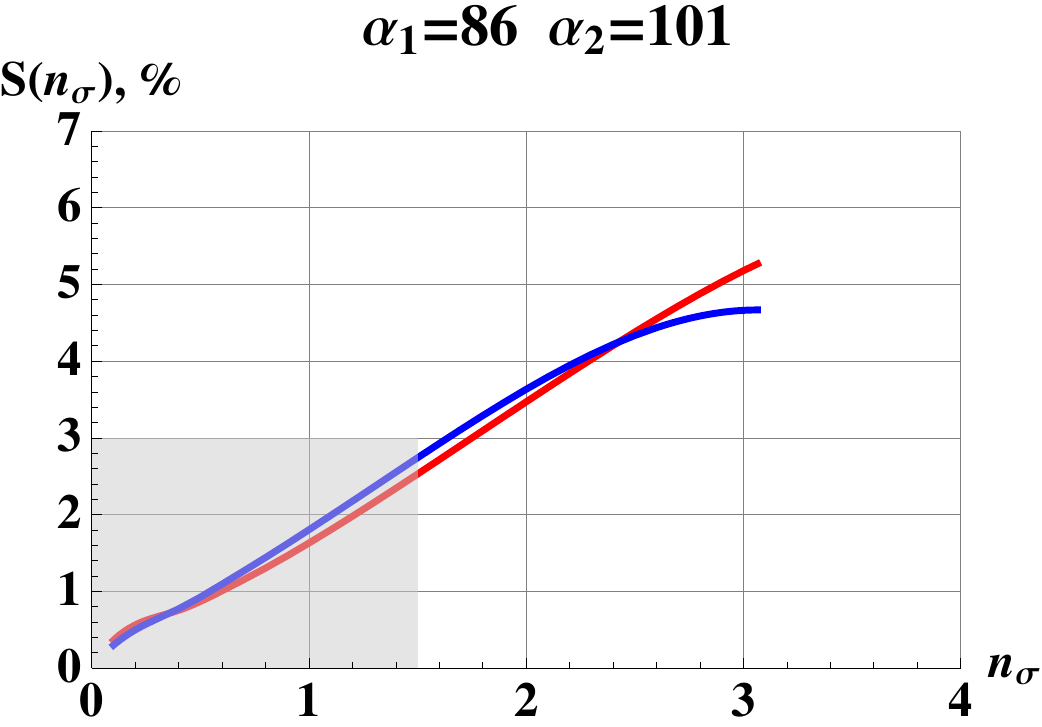}\\
  \includegraphics[width=.24\textwidth,angle=0]{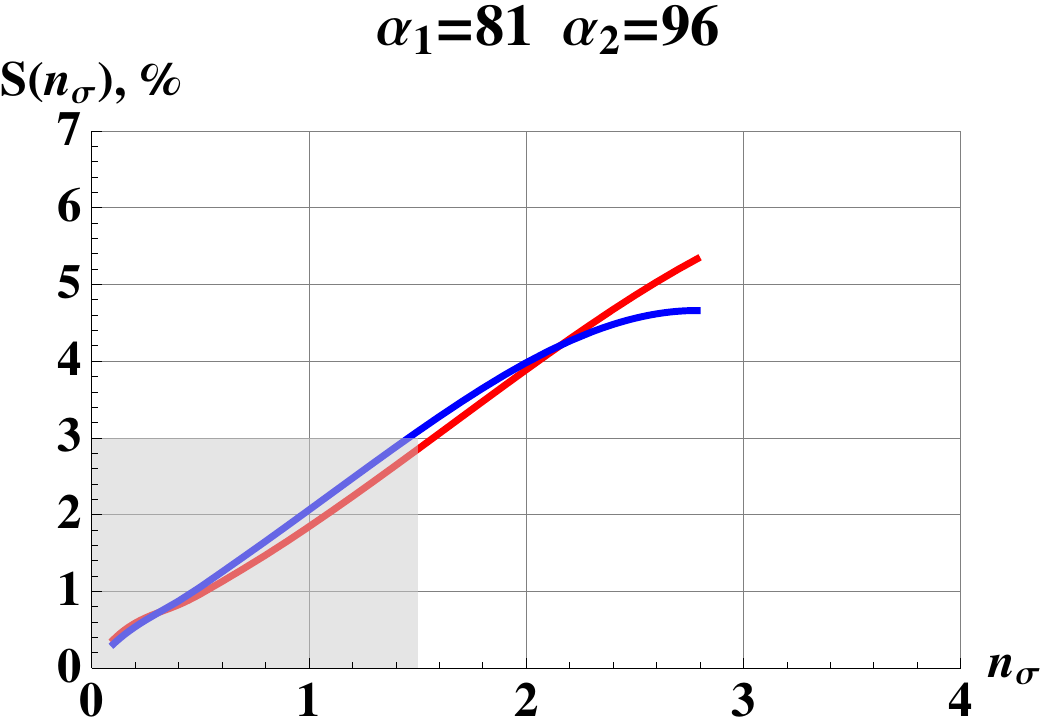}
  \includegraphics[width=.24\textwidth,angle=0]{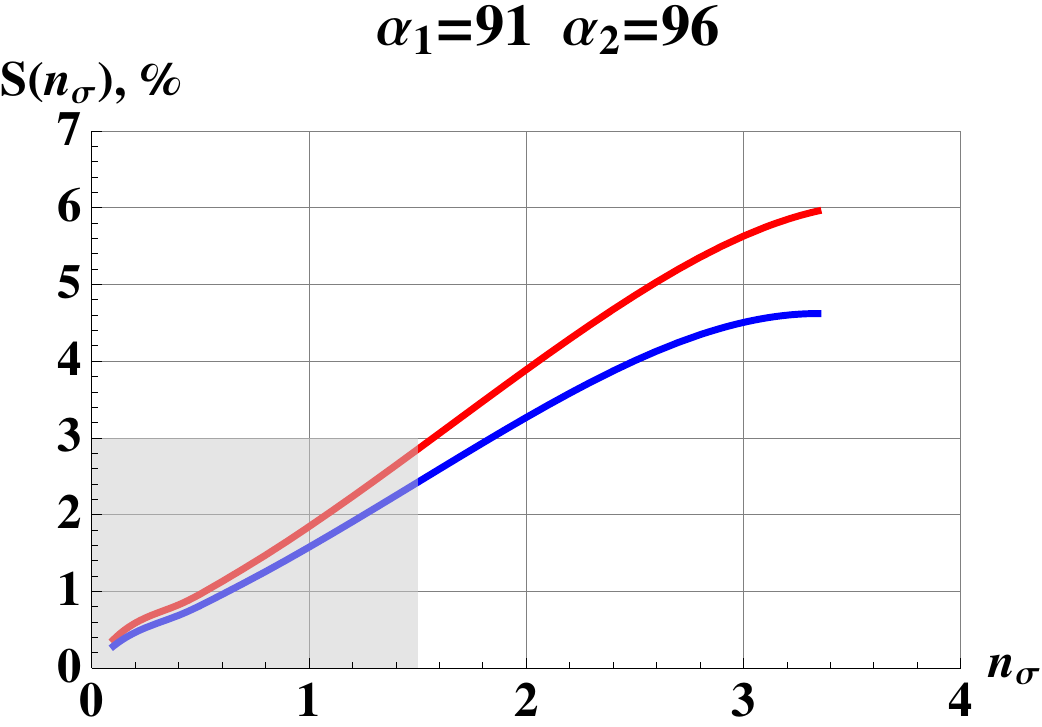}
  \caption{Small variations about the solution $\pm 5~\murad$.}
  \label{vary}
\end{figure}

In view of our previous findings, the off-plane losses (in IR5) are
neglected and by using the postulate made in the Introduction
(Eq.~\ref{params}), we have:
\begin{eqnarray}
  S(n_{\sigma};1.2\times10^{11}, \alpha_1)  = S(n_{\sigma};1.6\times10^{11}, \alpha_2),
  \label{cond}
\end{eqnarray}
which is to be solved for the angles.
\begin{figure}[H]\begin{center}
    \includegraphics[width=.35\textwidth,angle=0]{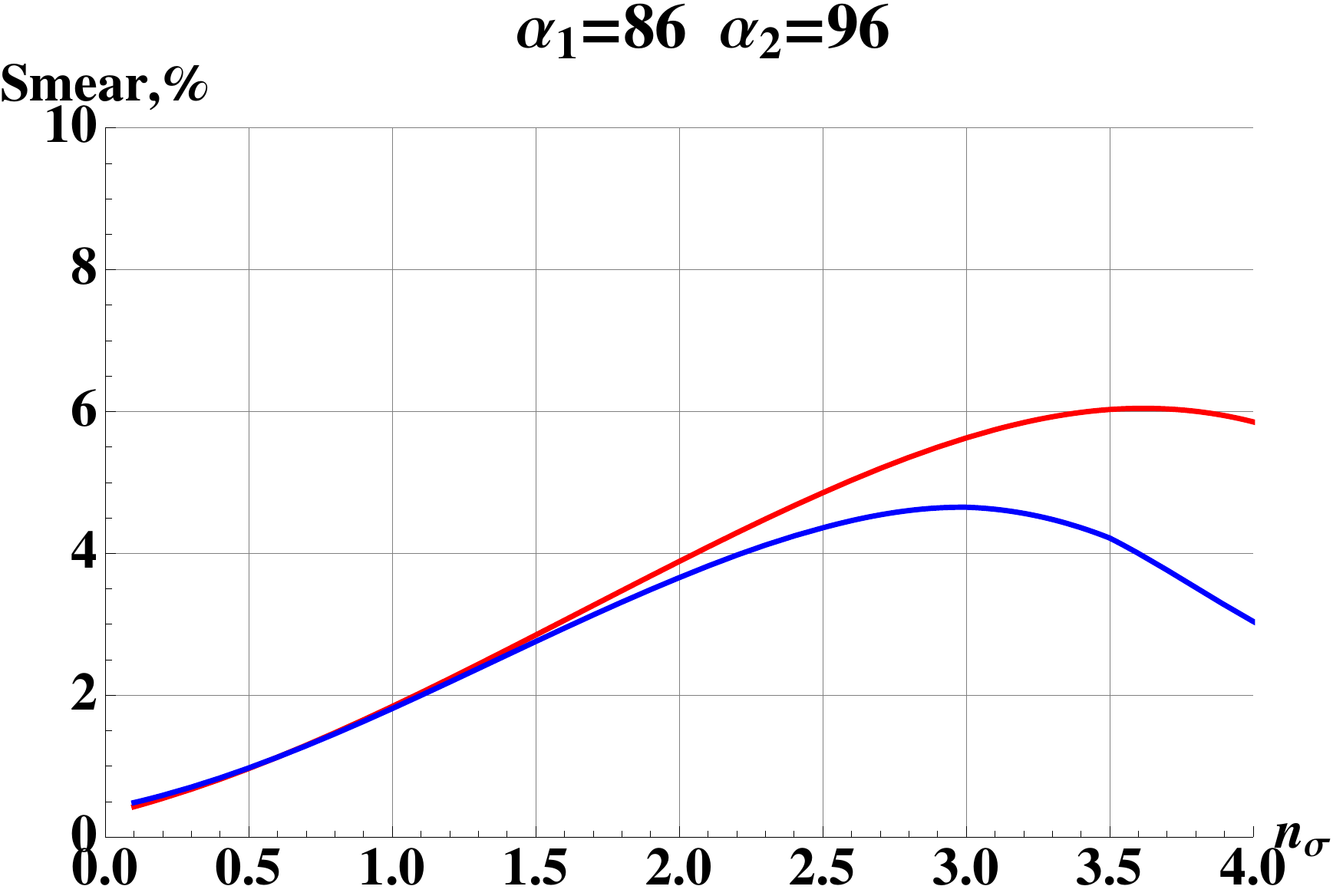}\end{center}
  \caption{Graphs of {\color{red} $S(n_{\sigma};1.2\times10^{11},86)$ }
    (red) and {\color{blue} $S(n_{\sigma};1.6\times10^{11},96)$ } (blue).
    The smear is seen to be $\approx 3\%$ at 1.5 $\sigma$.}
  \label{8496}
\end{figure}
Figure~\ref{8496} shows that a good solution to Eq.~\re{cond} consists of the
values $\alpha_1 = 86$, $\alpha_2 = 96 \murad$. Indeed, this figure shows
that Eq.~\re{cond} is fulfilled not in a single point, but for all
amplitudes up to 1.5 $\sigma$, where the smear reaches $\approx
3\%$.  What has happened, of course, is that scaling by a factor
$1.6/1.2$, but reducing the angle from $\alpha_2$ to $\alpha_1$, has
almost exactly preserved one particular blue branch from Fig.
\ref{deponang}.  Conversely, small variations about this solution, say
$\pm 5~\murad$, lead to deviations of red and blue curves, as shown in
Fig.~\ref{vary}.
\subsection{Explanation of Cases $N_{\lr}$=24 and 16 (Green and Black)
}
\begin{figure}[h]
  \includegraphics[width=.24\textwidth,angle=0]{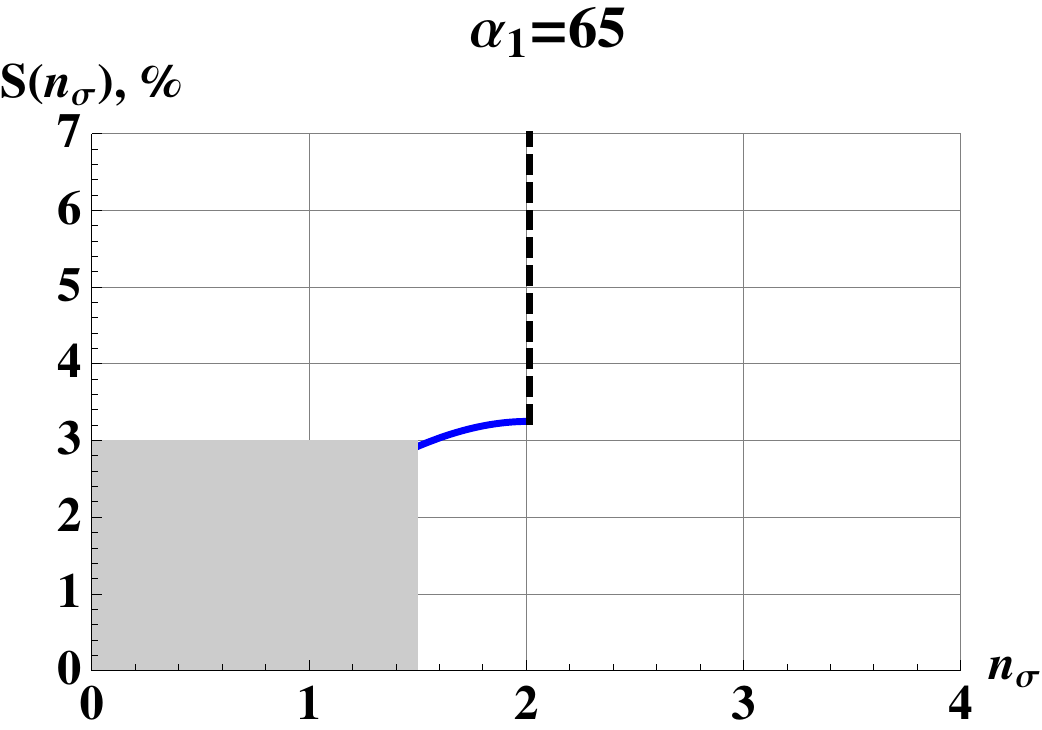}
  \includegraphics[width=.24\textwidth,angle=0]{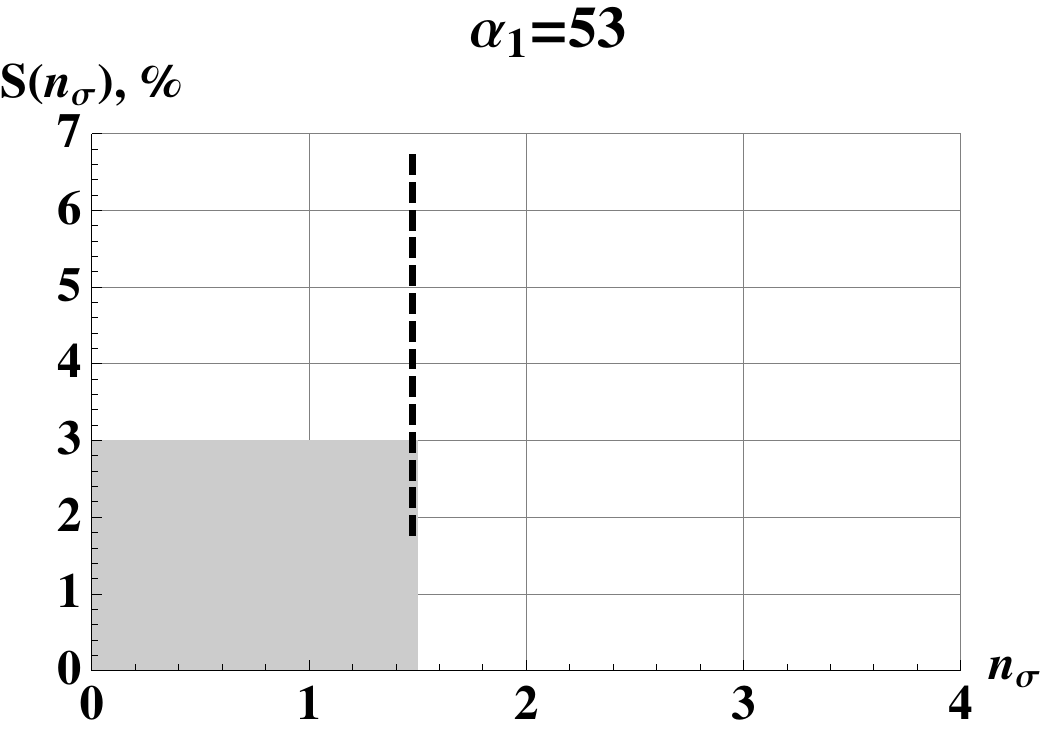}
  \caption{$\np=1.2\times10^{11}$.}
  \label{crang12}
\end{figure}

\begin{figure}[h]
  \includegraphics[width=.24\textwidth,angle=0]{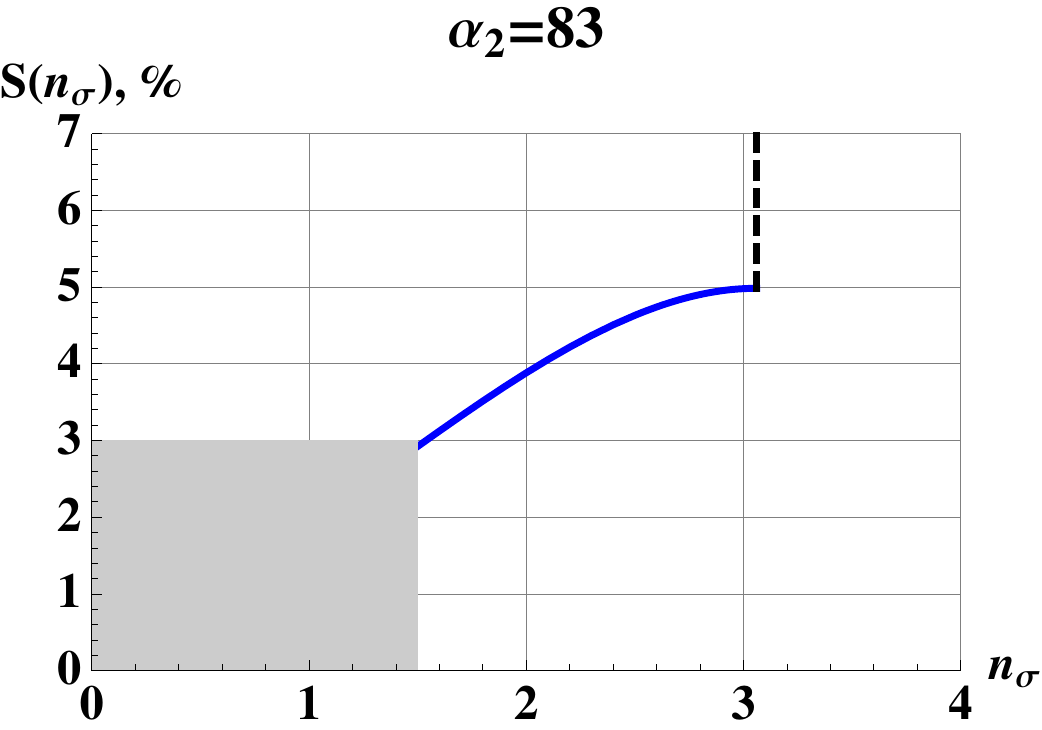}
  \includegraphics[width=.24\textwidth,angle=0]{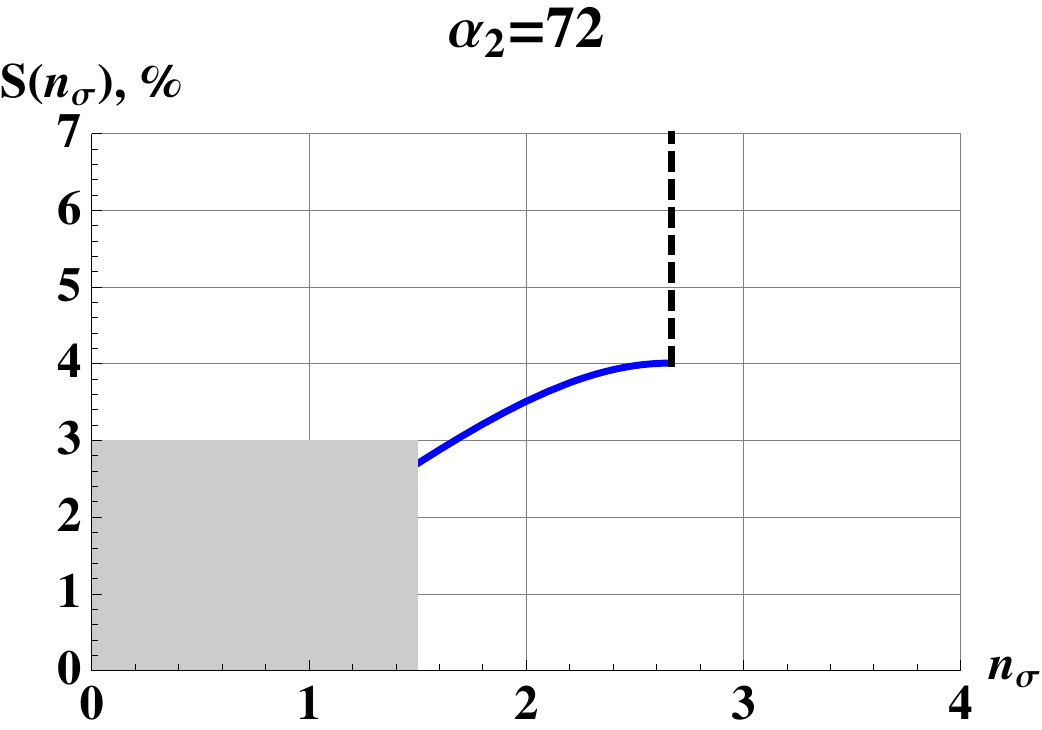}
  \caption{$\np=1.6\times10^{11}$.}
  \label{crang16}
\end{figure}

For $N_{\lr}$=24 and 16 (reduced collision sets in Fig.~\ref{drops}),
one needs to explain the green and black decay curves in
Figs.~\ref{exp2} and \ref{exp1}. By looking now at the bottom two plots
in Fig.~\ref{deponang}, we search for blue branches that pass through
the same maximum-smear point as found above: $3\%$ at 1.5
$\sigma$. The resultant branches are plotted in Figs.~\ref{crang12}
and \ref{crang16}, with solution angles as summarized in Table~1. 
Again,
at least a qualitative agreement is observed to the extent allowed by
the resolution of Figs.~\ref{exp2} and \ref{exp1}.
\begin{table}[H]
\caption{Angles of solutions for different intensities.}
   \label{tab:01}
  \begin{center}
    \begin{tabular}{ccc}
    \toprule
      $\np$      &    Green     &Black \\
    \toprule
      $1.2\times10^{11}$  &     65 & 53 \\
    \midrule
      $1.6\times10^{11}$    & 83 & 72 \\
    \bottomrule
    \end{tabular}
  \end{center}
\end{table}
Of the four plots in Figs.~\ref{crang12} and \ref{crang16}, on three
occasions the 3\%-smear line intersects a monotonic part of $S(\en)$
where, as we already know from Section 4, there is an exact agreement
with SixTrack. The rough indication for the dynamic aperture, as the amplitude
corresponding to a maximum of $S$, has been used in only one case:
$\alpha$=53.

\section{ Appendix } 
\appendix
The non-linear kick factor in Eq.~\re{fnon} is 
\begin{eqnarray*}
  &\Fnon(\en,\phi) =  \gamma+ \Gamma_0(P) +\mathrm{ln}(P)
  -F_{(1)} -F_{(2)} \,,\\
  &P
  = \frac{1}{2} \left( ( n_x+ \en \sin{\phi})^2 + n_y^2  \right)\,,
  \\
  &F_{(1)}
  = \frac{ 2 n_x }{  ( n_x^2+ n_y^2) } \left( 1-\mathrm{e}^{- \frac{n_x^2+ n_y^2 } {2 } } \right)  \en  \sin{\phi} \,,
  \no \\
  &F_{(2)}
  =\frac{
    -n_x^2+n_y^2+\mathrm{e}^{-\frac{n_x^2}{2}-\frac{n_y^2}{2}}
    \left(n_x^2+n_x^4-n_y^2+n_x^2 n_y^2\right)}{
    \left(n_x^2+n_y^2\right)^2 }\times \\
  &\times\, \en^2\sin^2\phi.  \,\ \ \ \ \ \ \ \ \ \ \ \ \ \ \ \ \  \ \ \ \ \ \ \ \ \ \   \ \ \ {\rm (A.1)}
\end{eqnarray*}
By following Ref.~\cite{irwinsmear}, the Lie map is given by an expression
of the following form:
\begin{eqnarray*}
  &  M_{N+1 } \e{f^{(N)}}   M_{N} \dots \e{f^{(2)}}
  M_2 \e{f^{(1)}} M_{1} \,,\no \\
  &f^{(k)}(x)\equiv \Fnon^{(k)}(x).
\end{eqnarray*}
Here, $M_k $ are linear operators and for brevity we have replaced
$\Fnon^{(k)}(x)$ with $f^{(k)}(x)$ .  We will show that since $\Fnon$
depends only on the normalized coordinate $x/\sigma$, once we rewrite
it in terms of the eigen-coordinates at the $k$th kick, the local
beta functions $\beta^{(k)}$ disappear, while the phase $\phi^{(k)}$ is
simply added to $\phi$.

By reversing the order, the map transforming the test particle
$(x,p_x)$ for one turn around the ring is
\begin{eqnarray*}
  & {\cal M}= M_{1} \e{f^{(1)}}   M_{2} \e{f^{(2)}} \dots
  M_N \e{f^{(N)}} M_{N+1} =\\
  &= \e{\overline M_1 f^{(1)}} \e{ \overline M_{2} f^{(2)}}  \dots \e{\overline M_N f^{(N)}} \overline  M_{N+1}.
\end{eqnarray*}
Reversal of the order means that in the first line all $ f^{(k)}$ are
now functions of the same initial variables $(x,p_x)$.  In the second
line, accumulated linear maps $\overline M_k = M_1 M_2 ... M_k$ have
been applied to transform the initial vector to the kick location.
Thus, as a first step, we have moved all kicks to the front of the
lattice and $\overline M_{N+1}$ is the total one-turn linear Lie
operator.

Let us denote matrices corresponding to Lie operators with hats; for example,
$\widehat M_{N+1}$.  As a second step, with $\beta$, $\alpha$
being matched Twiss parameters at the end of the lattice, one uses an
$A_0$ transform that transforms the ring matrix to a rotation
(inserting identities ${\cal A}_0 {\cal A}^{-1}_0 $ in between the
exponents):
\begin{eqnarray*}
  \widehat  M_{N+1} \stackrel{{\widehat{ \cal A}_0}}{\longrightarrow}
  {\footnotesize \widehat R= \left(
      \begin{array}{cc}
        \cos \mu & \sin \mu\no \\
        -\sin \mu& \cos\mu
      \end{array}
    \right) },
\end{eqnarray*}
\begin{eqnarray*}
  {\widehat{\cal A}_0} = 	\left(
    \begin{array}{cc}
      \sqrt\beta  & 0\\
      -\alpha/\sqrt{\beta}& 1/\sqrt{\beta}
    \end{array}
  \right).
\end{eqnarray*}
The two steps above combined are equivalent to replacing the argument
of $f$ by $\widetilde x_k$ -- the eigen-coordinate at the $k$th
location. To see this, apply the ${\cal A}_0$ transform to both
kick factor and coordinate:
\begin{equation*}
  {\cal A}_0 \overline M_{k} f^{(k)}(x) = f^{(k)}({\cal    A}_0
  \overline M_k x) = f^{(k)}(\widetilde x_k )\,,
  \label{ftilde1}
\end{equation*}
\begin{eqnarray*}
  &&  {\widetilde x_k}  \equiv {\cal A}_0  \overline M_k x = \sqrt{2\beta^{(k)} J } \sin{(\phi +\phi^{(k)})}\;.
\end{eqnarray*}
One can now drop the ${\cal A}_0$ on both sides of $\cal M$ and
consider the map:
\begin{eqnarray*}
  &&  {\cal M} = \e{\widetilde f^{(1)} } \e{\widetilde f^{(2)}} \dots
  \e{\widetilde f^{(\nip)}}   R, \\
  &&  \widetilde f^{(k)}(\jx,\phi)=f^{(k)}( \widetilde x_k), \\
  && R=\e{f_2} ,\ \ \ \ \ \ \ {:f_2:}=-\mu \jx .
\end{eqnarray*}
To first order, one can just sum the Lie factors:
\begin{eqnarray*}
  &   {\cal M}  \approx  \e{F}\;R =\e{h},\ \ \ \ \ \ \  F \equiv  \sum_{k=1}^{\nip} \widetilde f^{(k)}.
\end{eqnarray*}
By noting that above, as in Ref.~\cite{irwinsmear}, $R$ precedes the
kick, while in Eq.~\re{oneip} and Ref.~\cite{chaolect} the kick is assumed to
be at the end of the lattice, our map is identical to Eq.~\re{oneip}.

The first-order invariant $h$ is now found with the BCH theorem.  Let
us write $F=\bar F + F^\star$, where $F^\star$ is the oscillating part.
By taking only $F^\star$:
\begin{eqnarray*}
  &&  h(\jx,\phi)=   f_2  + \frac{:f_2:}{1-\mathrm{e}^{-:f_2:}} F^\star, \ \ \
  {\rm (A.2)}\\
  &&  F^\star \equiv  \sum_{k=1}^{\nip} (\widetilde f^{(k)})^{\star}, \no
  \label{hone}
\end{eqnarray*}
where according to Eq.~\re{ftilde},
\begin{eqnarray*}
  (\widetilde f^{(k)})^{\star}=  \sum_{m\neq 0} C_m^{(k)} \mathrm{e}^{i m \phi}
  = \sum_{m=1}^{\infty} \left(C_m^{(k)} \eplm + cc \right) \,.
\end{eqnarray*}
A basic property of ${:f_2:}$ is to operate in a simple way on
functions of $\jx$, or eigenvectors $\mathrm{e}^{i n \phi}$.  Also, functions
$G(f_2)$ can easily be applied to eigenvectors:
$${:f_2:}\; \mathrm{e}^{i n \phi} = i\, n\, \mu\, \mathrm{e}^{i n  \phi}, $$
$$G({:f_2:}) \mathrm{e}^{i n \phi} = G( i\, n\, \mu\,) \mathrm{e}^{i n  \phi}. $$
If we choose $G({ \colon f_2 \colon })\equiv \frac{:f_2:}{1-\mathrm{e}^{:f_2:}
}$, then we have:
\begin{eqnarray*}
  && : G(f_2) :  \eplm  = \\
  &&=G (i m   \mu ) \eplm =\\
  &&= \frac{i m   \mu }{1-\mathrm{e}^{-i m   \mu}} \eplm =\\
  && =   \frac{i\, m\, \mu\, \eplm }{\mathrm{e}^{i\, m\, \mu/2}  -\mathrm{e}^{-i\, m\, \mu/2}} \mathrm{e}^{i\, m\, \mu/2}  = \\
  &&= \frac{m\, \mu\, \eplm }{2\, \sin{ (m\, \mu/2)}} \mathrm{e}^{i\, m\, \mu/2}.
\end{eqnarray*}
%
By substituting all these in Eq. (A.2) and using the property $C_m^{(k)}=
c_m^{(k)} \mathrm{e}^{i m \phi^{(k)}}$, we obtain:
\begin{eqnarray*}
  && h(\jx,\phi)=\no \\
  &=& -\mu \jx - \displaystyle  \bb\; \sum_{k=1}^{N} \sum_{m=1}^{\infty} \left( \frac{m\; \mu\; c^{(k)}_m}{2      \sin{ (m \mu/2)}} \mathrm{e}^{i m (\mu/2 +\phi + \phi^{(k)})} + c.c. \right).
\end{eqnarray*}

\begin{thebibliography}{99}
\bibitem{footnote1}
The use of $\epsilon_n=2\times10^{-6}$
as an alternative does not change any results.
\end{thebibliography}
 \end{document}